\documentclass[review,10pt]{elsarticle}
\usepackage{amssymb}
\usepackage{amsthm}
\usepackage{graphicx}
\usepackage{epsfig}
\usepackage{latexsym, amsmath}
\usepackage[english]{babel}
\usepackage{amsmath, amstext, amsfonts}
\usepackage{float}
\usepackage{caption}
\usepackage{subcaption}
\usepackage{appendix}
\pdfoutput=1

\def\R{{\mathbf R}}
\def\Z{{\mathbf Z}}

\newtheorem{theorem}{Theorem}[section]

\newtheorem{Definition}[theorem]{Definition}

\journal{International Journal of Bifurcation and Chaos}

\begin{document}

\pagenumbering{arabic}
\begin{frontmatter}

\title{Multistability in Piecewise Linear Systems by Means of the Eigenspectra Variation and the Round Function}

\author[ipicyt]{H.E.~Gilardi-Vel\'azquez}
\ead{hector.gilardi@ipicyt.edu.mx}
\author[CARAO]{L.J.~Onta\~n\'on-Garc\'ia }
\ead{luis.ontanon@uaslp.mx}
\author[ipicyt]{D.G.~Hurtado-Rodriguez}
\ead{diana.hurtado@ipicyt.edu.mx}
\author[ipicyt]{E.~Campos-Cant\'on \corref{cor1}}
\ead{eric.campos@ipicyt.edu.mx}
\cortext[cor1]{Corresponding author}

\address[ipicyt]{
 \textsc{Divisi\'on de Matem\'aticas Aplicadas,\\}
 {Instituto Potosino de Investigaci\'on Cient\'{\i}fica y Tecnol\'ogica A.C.}\\
 \textsc{Camino a la Presa San Jos\'e 2055 col. Lomas 4a Secci\'on, 78216,\\
 San Luis Potos\'{\i}, SLP, M\'{e}xico \vskip 2ex}}

\address[CARAO]{
  {\textsc{Coordinaci\'on Acad\'emica Regi\'on Altiplano Oeste},\\}
  {Universidad Aut\'{o}noma de San Luis Potos\'{\i}},\\
  \textsc{Kilometro 1 carretera a Santo Domingo, 78600,\\
  Salinas de Hidalgo, San Luis Potos\'{\i}, M\'exico
  \vskip 2ex}}

\setcounter{page}{1}

\begin{abstract}
A multistable system generated by a Piecewise Linear  (PWL) system based on the jerky equation is presented. The systems behaviour is characterised by means of the Nearest Integer or $round(x)$ function to control the switching events and to locate the corresponding equilibria among each of the commutation surfaces.
These surfaces are generated by means of the switching function  dividing the space in regions equally distributed along one axis.
The trajectory of this type of system  is governed by the eigenspectra of the coefficient matrix which can be adjusted by means of a bifurcation parameter. The behaviour of the system can change from  multi-scroll attractors into a mono-stable state to the coexistence of several single-scroll attractors into a multi-stable state. Numerical results of the dynamics and bifurcation analyses of their parameters are displayed to depict the multi-stable states.

\end{abstract}


\end{frontmatter}

\section{\label{sec:Introduction}Introduction}

Throughout history, the scientific community has had the task of studying the properties of different dynamical systems and their importance to the environment where they are developed. For example, the fact that some dynamical systems are chaotic and have a critical dependence on initial conditions is known since late last century. This important characteristic  has been correlated with the coexistence of several possible final stable states for a given set of parameters \cite{ulrike,Pisarchik,Sharma,Kapitaniak,Arecchi}. In dissipative systems, this property is called  ``generalized multistability" so it can be distinguished from the ordinary coexistence of stationary solutions \cite{Sharma,Arecchi}.

The term multistability was first used with respect to visual perception in 1971 \cite{Attneave}; nevertheless, the occurrence of multistability is very common in various fields of science, such as chemistry \cite{Showalter,Marmillot}, optics \cite{Saucedo, Brambilla}, physics \cite{Cai,Santer} and biological systems \cite{Ozbudak, Zhusubaliyev}. Nevertheless, the importance of generating multistable structures resides in the wide variety of applications that exist: synchronization, complex networks, communication, climate, etc.

In this work it is taken advantage of the properties of the hybrid dynamical systems \cite{haddad,aihara},  such as the Unstable Dissipative Systems  theory based on Piecewise Linear (PWL) systems whose solution presents multi-scroll attractors \cite{campos2}. The method proposed here consists on designing systems taken from the jerky equation which present the coexistence of  multistable state if the following conditions are met: i) if the distance between the commutation law and the equilibrium point satisfies a specific ratio, and ii) if the commutation law changes automatically regarding the values given to the Nearest Integer Function or $round(x)$, dividing the space in equally distributed subdomains given along an axis depending on the initial state applied to the system. Although there have been several reports on how to generate multistability, this is the first in our knowledge in which the values
of the eigenspectra presented in the systems are being affected directly by means of the $round(x)$ function. It will be shown that regarding these values, the system changes from multi-scroll attractors into a mono-stable state to the coexistence of several single-scroll attractors into a multistable state.

This paper is organized as follows: In section 2, the unstable dissipative system theory is introduced to generate scrolls along the  $x_1$  axis and the function for the commutation surface displacement. In section 3, the  Nearest Integer or $round(x)$ function is analyzed as a basis for the commutation surface and equilibria displacement.  Section 4 contains the results about the transition from multi-scroll attractors to multistability. In order to explain the behavior of the system, bifurcation analysis of the parameters involved are studied along with the variation of the eigenspectra of the system. The results are displayed numerical depicting the multistable structures obtained. Finally conclusions are drawn in Section 5.

\section{\label{sec:UDS}UDS theory}

 Consider the dynamical systems which are defined by a class of  linear affine system given by:

\begin{equation}\label{ec_sistemA}
\dot{\mathbf{X}}=\mathbf{A}\mathbf{X} + \mathbf{B},
\end{equation}

\noindent
where $\mathbf{X}=(x_1,x_2,x_3)^T \in {\R}^3$ is the state vector,
$\mathbf{B}~=~(b_1,b_2,b_3)^T\in \R^3$ stands for a real  vector and
$\mathbf{A}\in \R^{3\times 3}$ denotes a linear operator that is not singular with entries $(a_{ij})$, $i,j=1,2,3$.
The equilibrium point of the system results in $\mathbf{X}^*=-\mathbf{A}^{-1}\mathbf{B}$.
The class of linear affine systems considered here, are those that present oscillations around the equilibria due to the stable and unstable manifolds $E^s$ and $E^u$, respectively. These manifolds are defined in a way that $\vartheta = (\vartheta_{1,2,3})$ is a set of column eigenvectors such that $\mathbf{A}\vartheta_i = \lambda_i\vartheta_i$ with $i = 1,2,3$;  $E^s = Span \{ \vartheta_1\}$ and $E^u = Span\{ \vartheta_{2,3}\}$. In order to present such oscillations and following a similar mechanisms as in \cite{campos2,UDS4,campos_udsII}, two types of dissipative systems with unstable dynamics have been studied which will be called unstable dissipative systems (UDS), however only one type of both will be considered here. This type is defined in the following way:

\begin{Definition}\label{def_USD3}
   A linear system $\dot{\mathbf{X}}=\mathbf{A}\mathbf{X}$ , where $\mathbf{X} \in {\R}^3$ is the state vector, $\mathbf{A}\in \R^{3\times 3}$ is a linear operator and $\lambda _i, i=1,2,3$, are the eigenvalues of $\mathbf{A}$. If  $\sum_{i=1}^3\lambda_i<0$, and  one $\lambda_1$ is real negative $ \lambda_1 <0 $, and two $\lambda_{2,3}$ are complex conjugated with positive real part $(Re\{\lambda_{2,3}\}>0)$  then the linear system will be called a UDS of the type I.
\end{Definition}

If the linear affine system given by  eq. (\ref{ec_sistemA}) satisfies the Definition \ref{def_USD3} with $\mathbf{B}=0$ then it is possible to generate an attractor $\mathfrak{A}$ by means of  a PWL system under the following considerations for the vector $\mathbf{B}$:

\begin{equation}\label{ec_sistemABu}
\begin{array}{c}
\dot{\mathbf{X}}=\mathbf{A}\mathbf{\mathbf{X}} + \mathbf{B}(\mathbf{X}),\\\\
\mathbf{B}(\mathbf{X})=\left\{
          \begin{array}{ll}
            \mathbf{B}_1, & \hbox{if $\mathbf{X}\in \mathcal{D}_1$;} \\
            \mathbf{B}_2, & \hbox{if $\mathbf{X}\in \mathcal{D}_2$;} \\
            \vdots & \vdots     \\
            \mathbf{B}_k, & \hbox{if $\mathbf{X}\in \mathcal{D}_k$.}
          \end{array}
        \right.
\end{array}
\end{equation}

The affine vector $\mathbf{B}$ must be a  switching function that changes depending on which domain $\mathcal{D}_i\subset \R^3$ with $\R^3=\cup_{i=1}^k\mathcal{D}_i$ the trajectory is located. The equilibria of system  \eqref{ec_sistemABu} are given by $\mathbf{X}^*_i=-\mathbf{A}^{-1}\mathbf{B}_i$, with $i=1,\ldots, k$,  and each vector $\mathbf{B}_i$ of the  system  is considered in order to generate a multiscroll attractor of system \eqref{ec_sistemABu}.

The idea of the method lies on defining vectors $\mathbf{B}_i$ in order to assure stability of a class of dynamical systems in $\R^3$ with oscillations within the attractor $\mathfrak{A}$. In a way that for any initial condition $\mathbf{X}_0 \in \mathfrak{B} \subset \R^3$, where $\mathfrak{B}$ is the basin of attraction, the system given by eq. \eqref{ec_sistemABu} induces in the phase space $\R^n$ the  flow $(\phi^t)_{t\in\R}$. Thus, each initial condition $\mathbf{X}_0 \in \mathfrak{B}$ generates a trajectory given by $\phi^t(\mathbf{X}_0): t\geq 0$ which is trapped in an attractor $\mathfrak{A}$ after  defining at least two vectors $\mathbf{B}_1$ and $\mathbf{B}_2$, as it is described in \cite{UDS4}.

This class of systems can display various multi-scroll attractors as a result of a combination of several unstable ``one-spiral'' trajectories. The number of scrolls depends on the number of vectors $B_i$, $i=1,\ldots, k$ introduced in the system, thus the equilibrium points are given by $\mathbf{X}^*_i = -\mathbf{A}^{-1}\mathbf{B}_i$  and  which trajectories oscillate around them. This is a consequence of an important feature of this kind of UDS, where they can result in one scroll attractor for each equilibrium point appropriately added in the domains $\mathcal{D}_i \subset \R^n$ in which the system is divided \cite{UDS4}.

\subsection{Generation of scrolls along $x_1$}

A convenient approach to build the matrix $\mathbf{A}$ and vector $\mathbf{B}$ is based on the linear ordinary differential equation (ODE) given by the jerky form: $\stackrel{...}{x}+a_{33}\ddot{x}+\dot{x}a_{32} +a_{31}x+\beta=0$.

The location where the scrolls are positioned can be understood from the following example which implies the coefficient matrix $\mathbf{A}$ from the jerky equation \cite{physic} and the affine vector $\mathbf{B}$ as follows:

\begin{equation}\label{ec_AB_R3}
  \mathbf{A}=\left(
    \begin{array}{ccc}
      0 & 1 & 0 \\
      0 & 0 & 1 \\
     -a_{31}&-a_{32}& -a_{33}
    \end{array}
  \right),
\mathbf{B}=\left(
    \begin{array}{c}
      0 \\
      0 \\
      b_3
    \end{array}
  \right).
\end{equation}

According with Definition \ref{def_USD3} the following entries of the matrix $\mathbf{A}$ will be considered:

\begin{equation}\label{ec_A}
  \mathbf{A}=\left(
    \begin{array}{ccc}
      0 & 1 & 0 \\
      0 & 0 & 1 \\
     -10.5&-7.0& -0.7
    \end{array}
  \right).
\end{equation}

And for the affine vector $\mathbf{B}$, the value of $b_3$ will commute according to the value of $x_1$ as follows:

\begin{equation}\label{ec_rule2}
    b_3(x_1)=\left\{%
\begin{array}{ll}
    c_1, & \hbox{if $x_1\geq x_{\pm i_{cs}}$;} \\
    c_0,   & \hbox{otherwise.} \\
\end{array}%
\right.
\end{equation}

Here $c_{0,1} \in \R$  determine the values of the equilibrium points since $\mathbf{X}^{*}_{i}=(b_3/a_{31},0,0)^T=(c_i/10.5,0,0)^T|i=0,1$, and $x_{\pm i_{cs}}$ stands for the location of the commutation surface given along the $x_1$ axis regarding the position that it takes with respect to the positive or negative axis and with $i\in \Z$.
Considering  $c_0=0$ and $c_1=0.9$ the equilibria are located at $\mathbf{X}^{*}_{0}=(0,0,0)^T$ and $\mathbf{X}^{*}_{1}=(0.6,0,0)^T$ displacing only along the positive $x_1$ axis. In order to consider an equally distributed scrolling around the equilibria,  the distance between the equilibrium points is calculated with the euclidean distance $\alpha(\mathbf{X}^{*}_{0},\mathbf{X}^{*}_{1})=\sqrt{({{x_1}^{*}_{0}}-{{x_1}^{*}_{1}})^2+({{x_2}^{*}_{0}}-{{x_2}^{*}_{1}})^2+({{x_3}^{*}_{0}}-{{x_3}^{*}_{1}})^2}$, resulting in $\alpha=0.6$. Therefore, the commutation surface that  generates two equally distributed scroll trajectories is given by  $x_{1_{cs}}=\alpha(\mathbf{X}^{*}_{0},\mathbf{X}^{*}_{1})/2$ resulting in a surface located at the $x_1=0.3$ plane. The eigenvalues and corresponding eigenvectors result in:

\begin{figure}[!t]
  \centering
\includegraphics[width=0.48\textwidth]{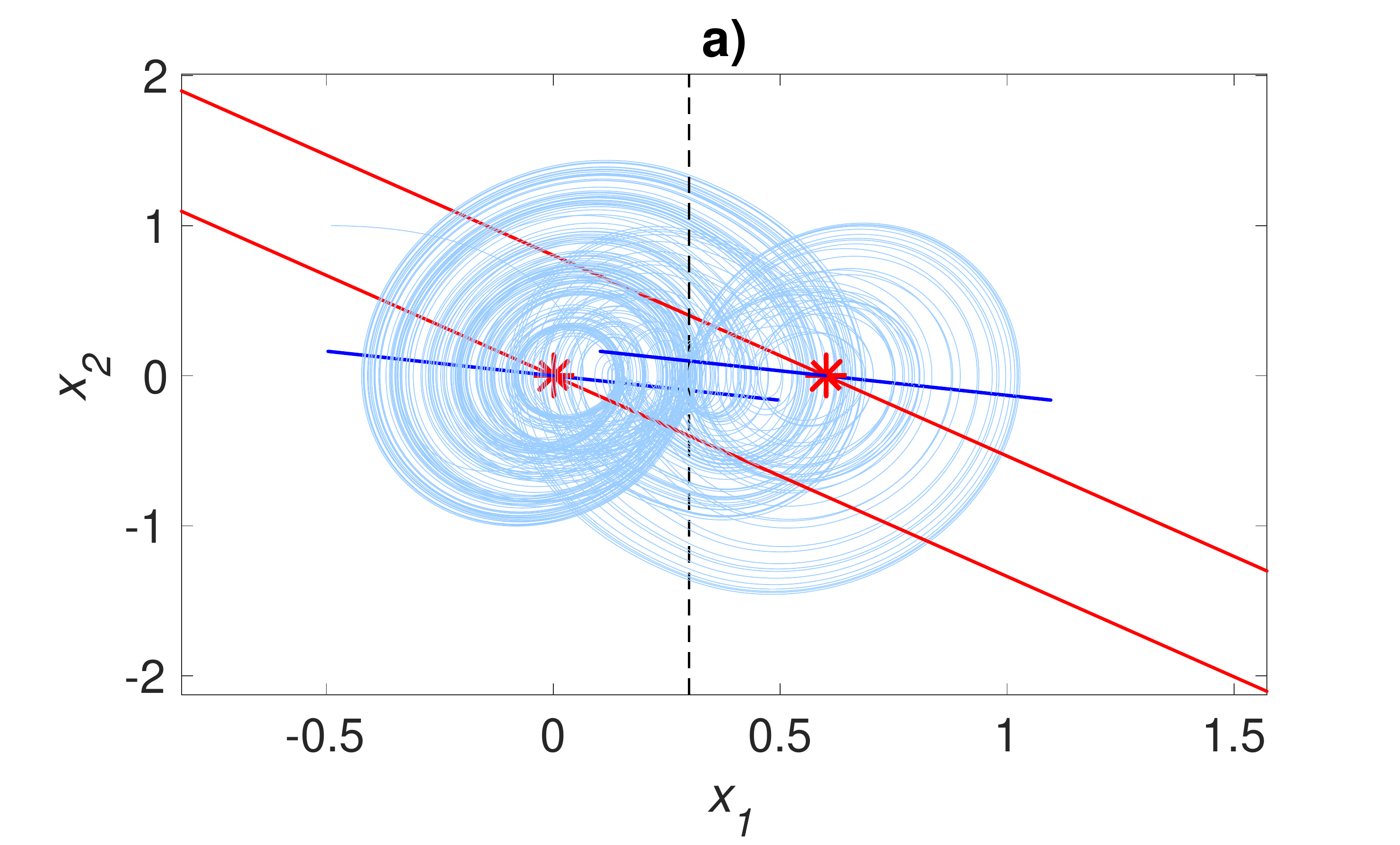}
\includegraphics[width=0.48\textwidth]{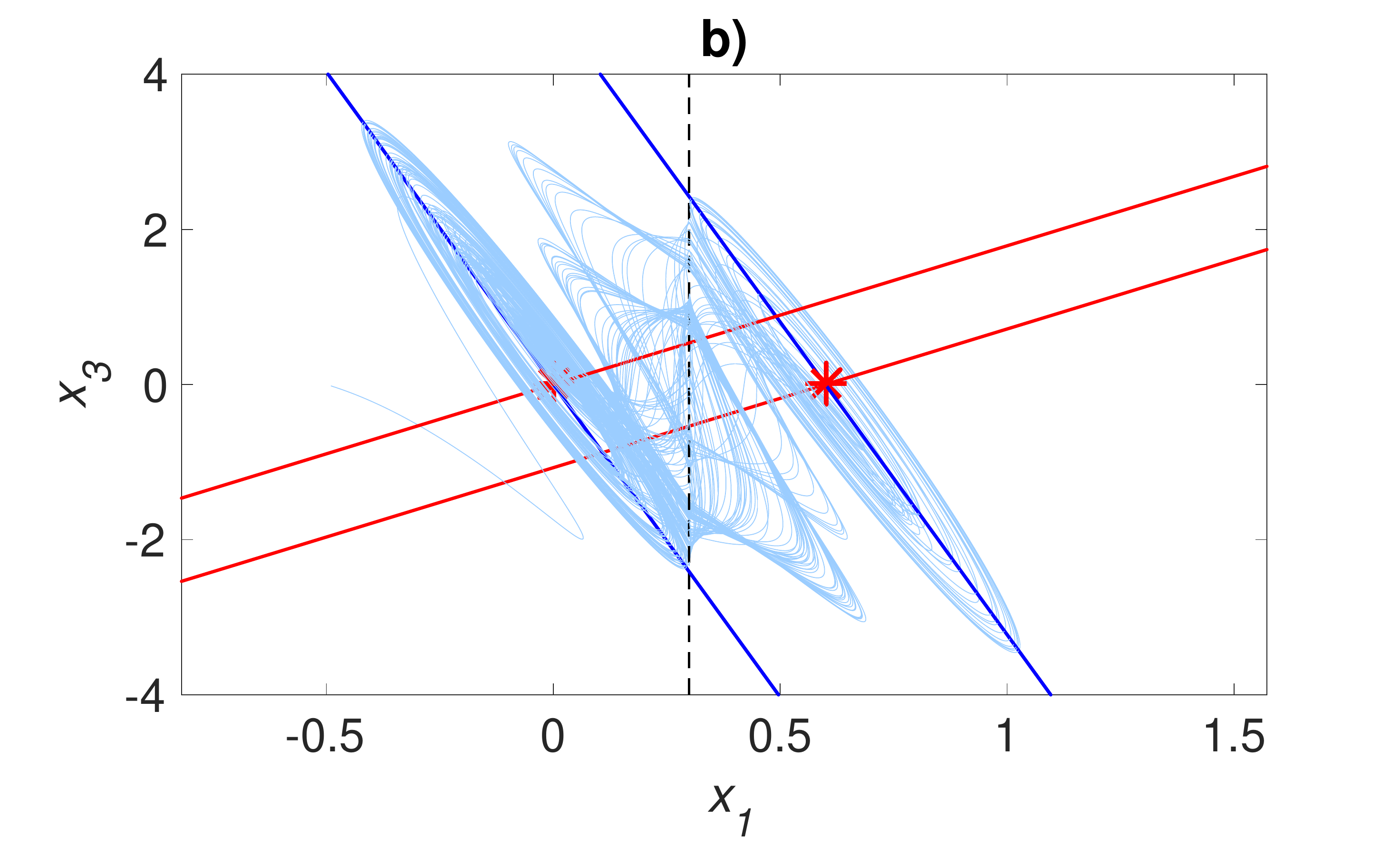}\\
\includegraphics[width=0.48\textwidth]{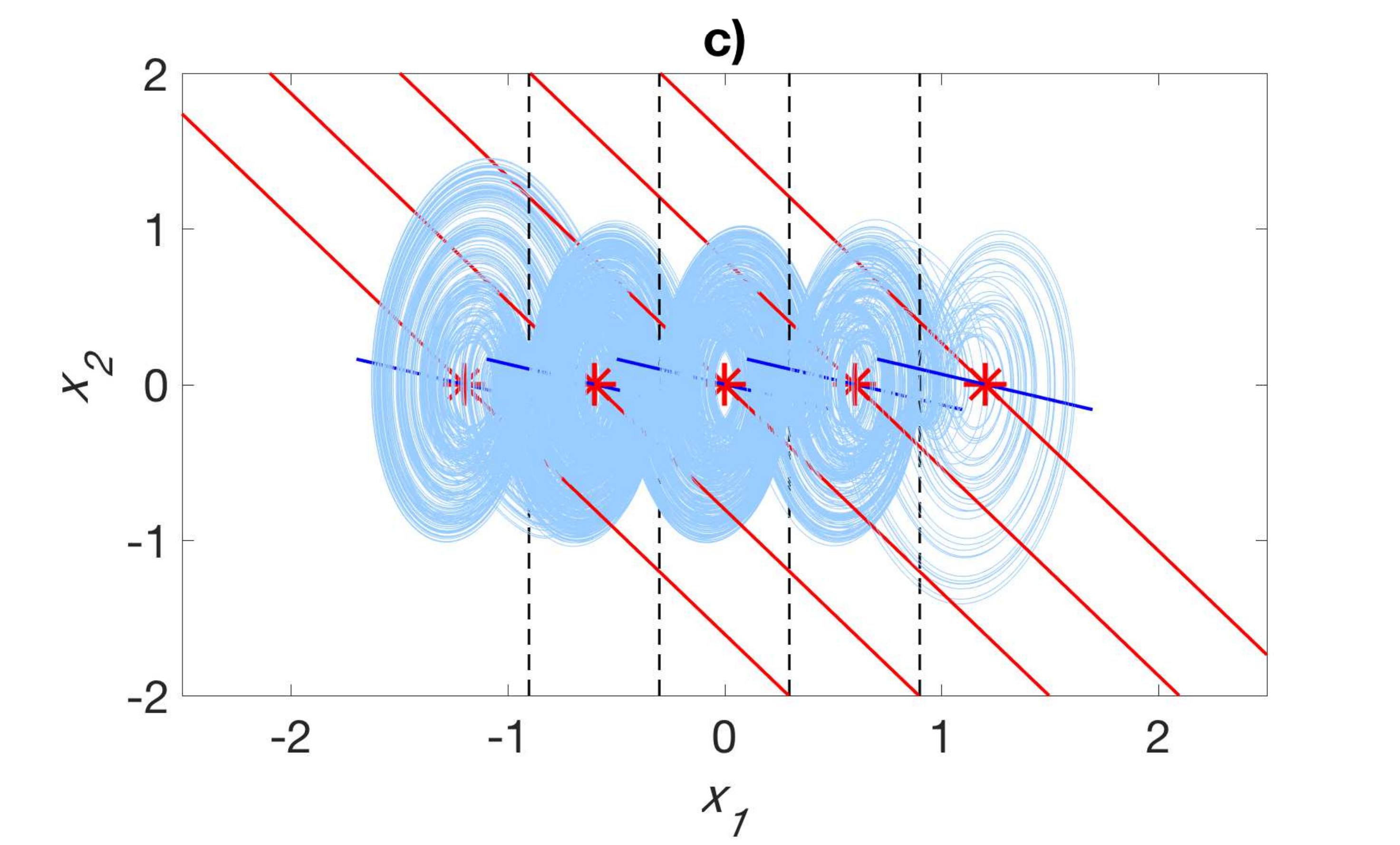}
\includegraphics[width=0.48\textwidth]{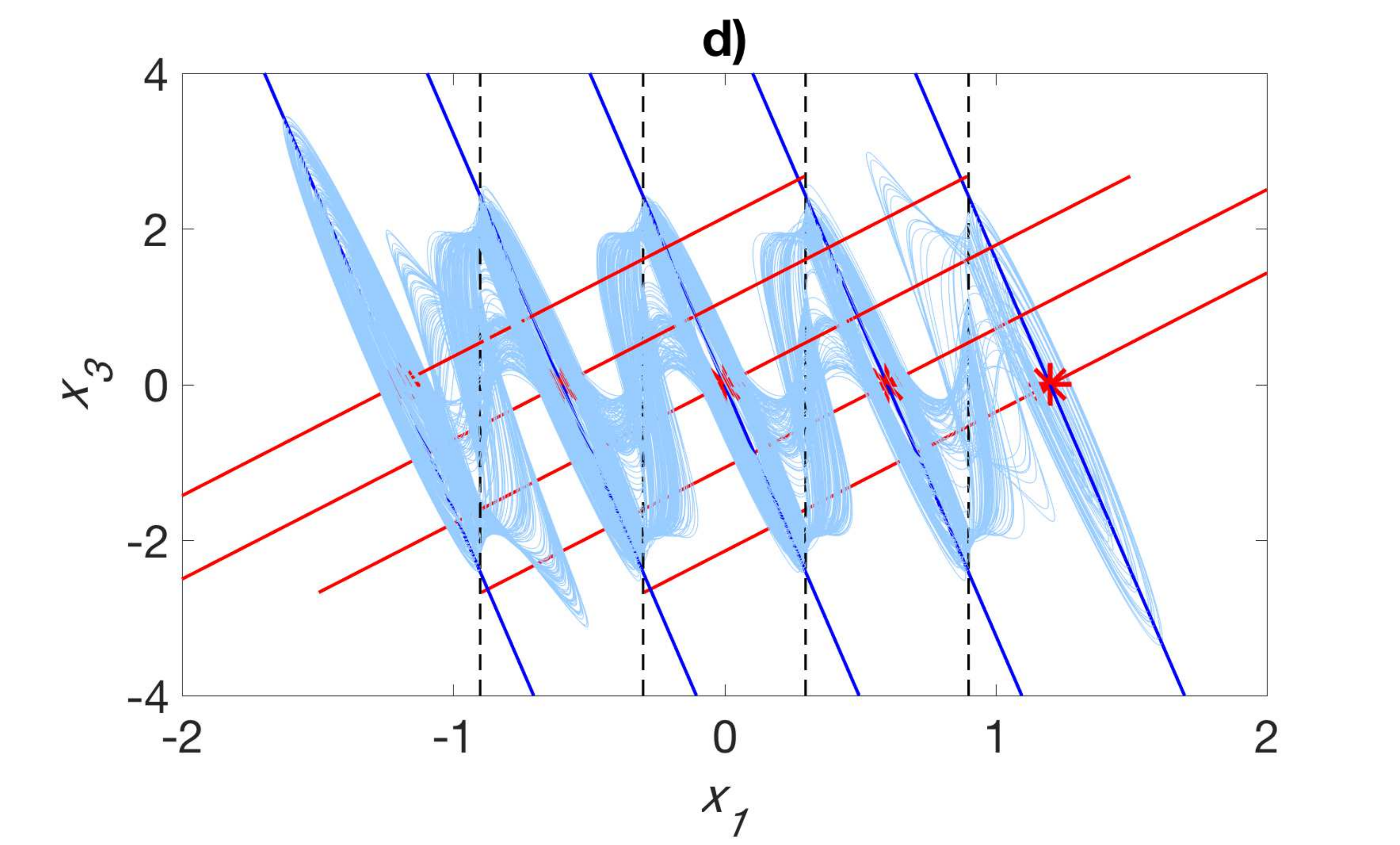}\\
\caption{\label{fig:Bi} Projections of the attractor given by eqs.  \eqref{ec_sistemABu} with \eqref{ec_A} and \eqref{ec_rule2} onto the $(x_1,x_2)$ plane for a) and c), and the $(x_1,x_3)$ plane for b) and d). The commutation law given by eq. \eqref{ec_rule2} for a) and b), and eq. \eqref{ec_rule6} for c) and d). Marked with red asterisks the equilibria of the system and with black dashed line the commutation surface. The direction of the eigenvectors are marked with red and blue lines for the stable and unstable manifolds $E^s$ and $E^{u}$, respectively.}
\end{figure}

\begin{equation}\label{ec_eig1}
\begin{array}{l}
    \lambda=\left\{-1.3372, 0.3186 \pm i 1.754\right\}.\\
    \mathbf{\vartheta}=\{\vartheta_{1,2,3}\} = \left\{\left(\begin{array}{c}0.4087 \\ -0.5466\\0.7309\end{array}\right)
\left(\begin{array}{c}-0.1160 \pm i 0.0269\\ 0.0379 \pm i0.3316\\0.9351\end{array}\right)
\right\}
\end{array}.%
\end{equation}%

{ The commutation surface must be located taking into consideration the manifolds $E^s, E^u$ } and between $\mathbf{X}^{*}_{0,1}$, otherwise the trajectory of the system can escape and the scrolls are not formed (this is described in Figure 2 and Figure 3 of the references \cite{campos2,UDS4}, respectively).  The trajectory of the system  given by the initial condition $\mathbf{X}_0=( 0.7,0,0)^T $ and eqs. \eqref{ec_sistemABu} with \eqref{ec_A} and \eqref{ec_rule2} presents a double scroll attractor as it is depicted in Figure \ref{fig:Bi} a) and b). Marked with red asterisks the equilibria of the system, and with red and blue lines the eigenvectors corresponding to the stable $E^s$ and the unstable $E^{u}$ manifolds ({\it i.e.} $E^s=Span\{\vartheta_1\}$, $E^u=Span\{Real(\vartheta_{2}),imag(\vartheta_{2})\}$), respectively. It is important to mention that the eigendirection determined by the two complex conjugate eigenvalues  has been represented only as the projection of a line, however this manifold corresponds to a plane, also these manifolds end at the commutation surface but they were projected to the other domains in order to clarify their directions with respect to the other manifolds.

\begin{figure}[!t]
  \centering
\includegraphics[width=11cm]{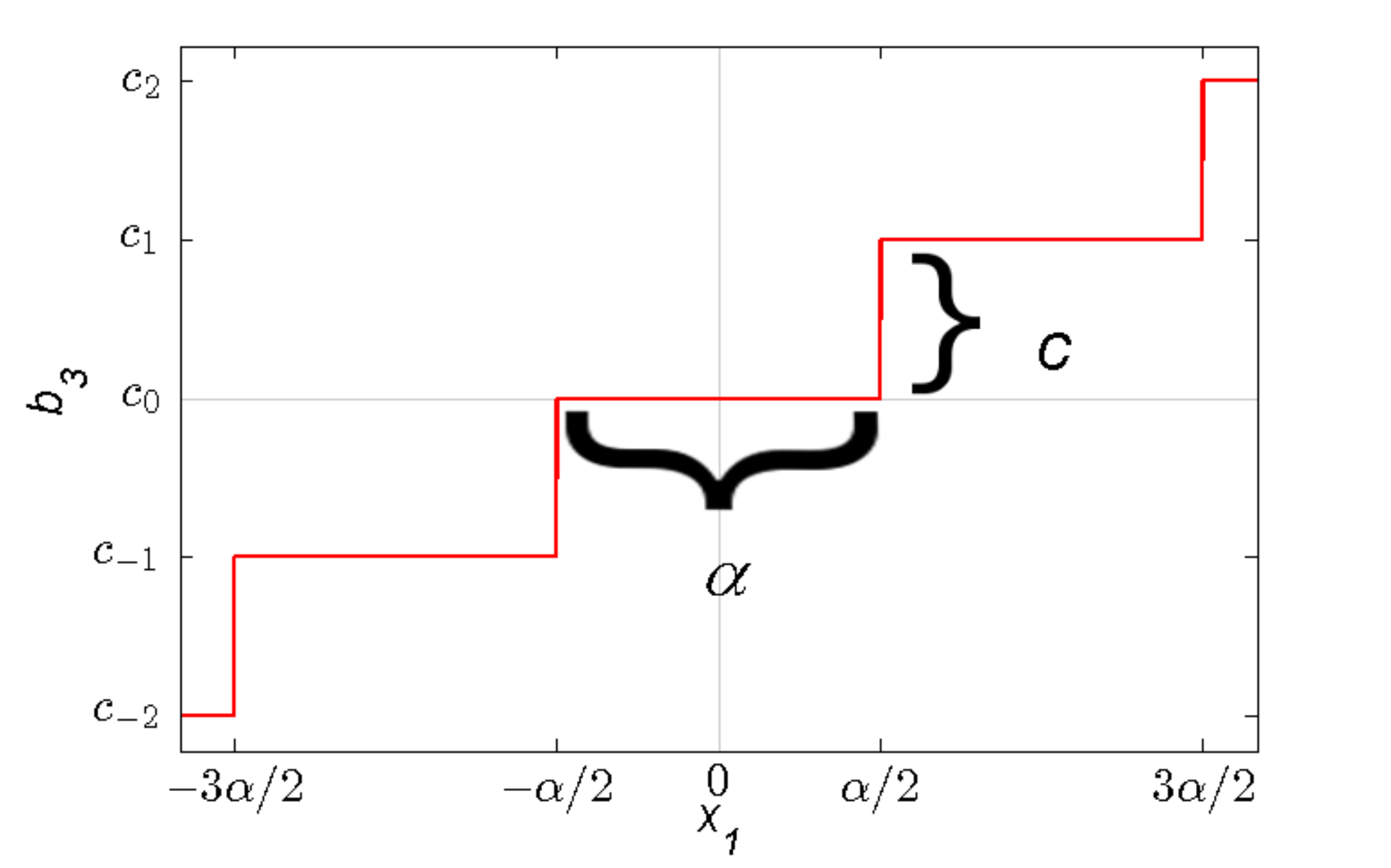}
\caption{\label{fig:round} Function $b_3$ given by  \eqref{ec_round}. The distance between each step is given by $\alpha$.}
\end{figure}

Between both scrolls, the commutation surface at the plane $x_1=0.3$ is marked with a black dashed line, dividing the space in two domains $\mathcal{D}_{0,1}$. Notice two important facts about the system, first that the scrolls are increasing their size due to the unstable manifold, this can be better appreciated at the projection of the attractor onto the  $(x_1,x_3)$ plane from Figure \ref{fig:Bi}  b). Second, that the trajectory of the system escapes from the domain $\mathcal{D}_{0}$ with $x_1<0.3$ located in the left side of the commutation surface. This occurs near the $E^{u}$ manifold where it crosses the surface in the lower part of the Figure \ref{fig:Bi}  b) and  it is attracted by $E^s$ towards the equilibrium point  in the domain $\mathcal{D}_{1}$ located at the right side of the commutation surface. The process is repeated in the inverse way forming scrolls around each equilibrium point.

This property of the UDS can  be easily extended for the generation of any number of scrolls along any of the axes if the above considerations are made when designing the system and locating  equilibria along the axes in the following way \cite{physcon,jimenez}.
First,  start considering $c_0=0$ (in case that the first equilibrium point is located at the origin) and the value of $c_1\neq 0$. Thus the commutation surface $x_{1_{cs}}$ that can result in two symmetric equilibrium points $\mathbf{X}^{*}_{i}, i=0,1,$ is given as $x_{1_{cs}}: x_1=\alpha(\mathbf{X}^{*}_{0},\mathbf{X}^{*}_{1})/2$, resulting in the two symmetrical scrolls. Now in order to extend it along the $x_1$ axis, the distance $\alpha(\mathbf{X}^{*}_{0},\mathbf{X}^{*}_{1})=0.6$ between the equilibria must be considered between adjacent equilibrium points in the system.

This idea of introducing more equilibria to the system can be done by considering the values $c_{\pm i}=\pm(a_{31}\alpha(\mathbf{X}^{*}_{0},\mathbf{X}^{*}_{1}) k)=\pm(10.5\alpha(\mathbf{X}^{*}_{0},\mathbf{X}^{*}_{1}) k)$, resulting in the consecutive equilibria along the $x_1$ axis $ \mathbf{X}^{*}_{\pm i} = (\pm \alpha(\mathbf{X}^{*}_{0},\mathbf{X}^{*}_{1}) k,0,0)^T $ with $i,k=0,\ldots,n$ ($i,k\in\Z^+$). The number of scrolls that are introduced in the 1D-grid along $x_1$  is  $2n+1$.
The commutation surfaces also should be located according to the value of the distance, in this case they are at $x_{\pm i_{cs}}=\pm \alpha(\mathbf{X}^{*}_{0},\mathbf{X}^{*}_{1})(1+2k)/2$. Therefore the commutation law for a $5$ scroll attractor is given by


\begin{figure}[!t]
  \centering
\includegraphics[width=11cm]{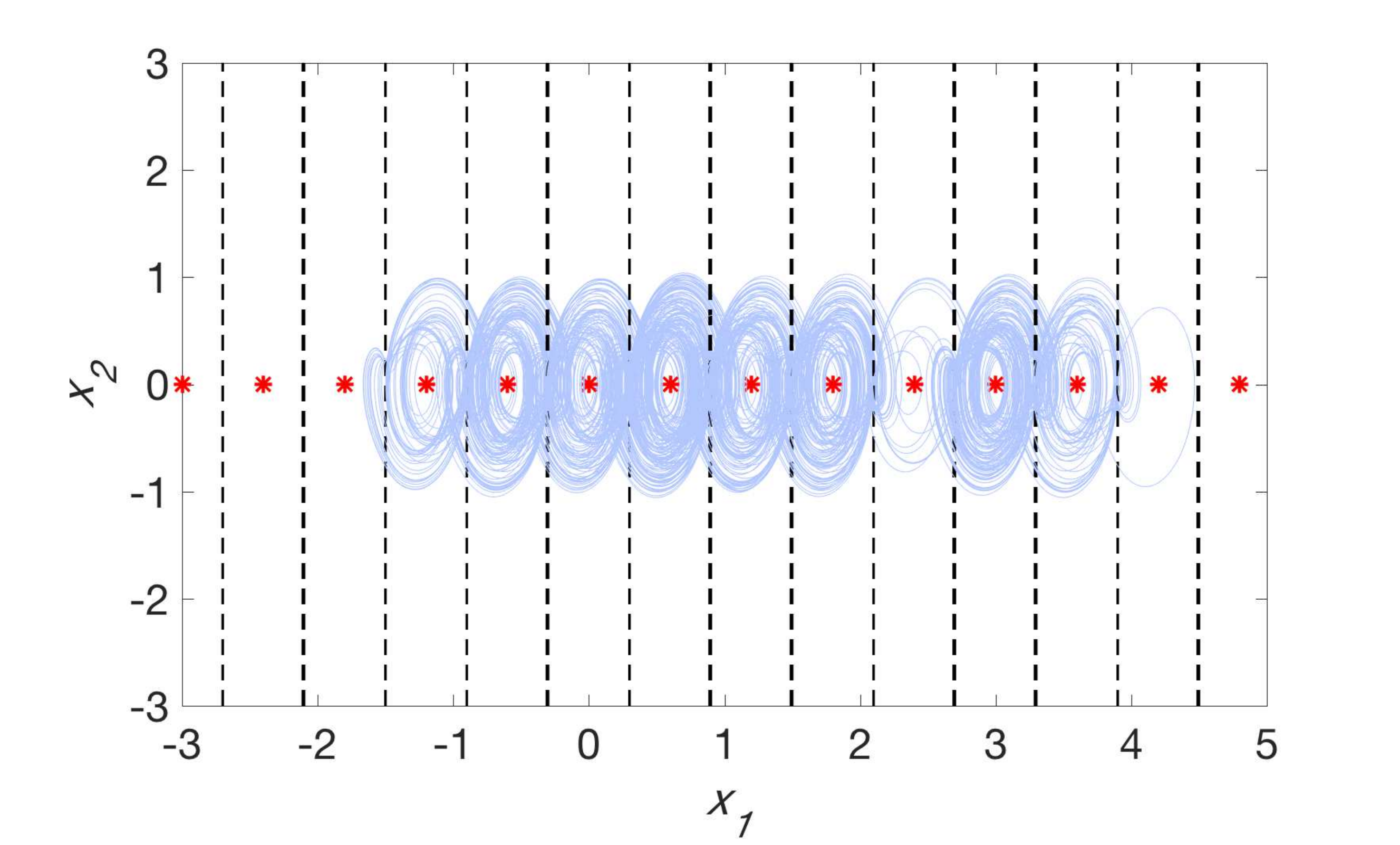}
\caption{\label{fig:atractor_round_1} Projection of the trajectory of the system \eqref{ec_sistemABu} given by eqs. \eqref{ec_AB_R3}  \eqref{ec_A} and \eqref{ec_round} onto the $(x_1,x_2)$ plane with $c=6.3$ and $\alpha=0.6$. Marked with red asterisk the equilibria of the system, and with gray line the commutation surfaces generated by the function  \eqref{ec_round}.}
\end{figure}

\begin{equation}\label{ec_rule6}
    b_3(x_1)=\left\{%
\begin{array}{ll}
    c_2, & \hbox{if $x_1\geq x_{2_{cs}}$ ;} \\
    c_1, & \hbox{if $x_{1_{cs}}\leq x_1<x_{2_{cs}}$ ;} \\
    c_0, & \hbox{if $x_{-1_{cs}}\leq x_1<x_{1_{cs}}$ ;} \\
    c_{-1}, & \hbox{if $ x_{-2_{cs}} < x_1\leq x_{-1_{cs}}$ ;} \\
    c_{-2}, & \hbox{if $x_1<x_{-2_{cs}}$ .} \\
  \end{array}%
\right.
\end{equation}

The resulting attractor  can be appreciated in Figures \ref{fig:Bi} c) and d). Both Figures present the commutation surfaces $x_{i_{cs}}$ marked with black line, along with the equilibria $\mathbf{X}^{*}_{\pm i}$. Each equilibrium point also depicts the $E^s$ and $E^u$ manifolds regarding their eigenvectors which are parallel among them similar as in the two scroll attractor.

\section{Function  for the commutation surface displacement}

Adding more equilibria to the system can  be easily implemented by using a step function instead of generating commutation surfaces manually, i.e., using an automatically step generating function as it has been applied in \cite{huerta}. Here, the $round(x)$ function will be implemented to simplify and automate this process. The function will be defined as follows:

\begin{equation}\label{ec_round_def}
    round(x)=\left\{%
\begin{array}{l}
    \lceil x+1/2 \rceil | 1/4(1+2x)\in \Z,\\
    \lfloor x-1/2 \rfloor | 1/4(-1+2x)\in \Z.
  \end{array}%
\right.
\end{equation}


For example, in order to have similar commutation surfaces and locations of equilibrium points that the ones described in eq. \eqref{ec_rule6}, consider the commutation of the vector $\mathbf{B}=(0,0,b_3)^T$ given by the following function:

\begin{equation}\label{ec_round}
    b_3(x_1)=c*round(x_1/\alpha),
\end{equation}

\noindent where $c\in \R$ corresponds to the amplitude of the function which is similar to the variable $c_i$, and $\alpha$ corresponds to the length of the step given by the round function centered in the origin, this is depicted in the graph of Figure \ref{fig:round}. Notice that the value of $\alpha$ has the same representation as $\alpha(\mathbf{X}^{*}_{0},\mathbf{X}^{*}_{1})$. If the function \eqref{ec_round} is considered as the commutation law of \eqref{ec_AB_R3}, then the commutation surfaces are located at every change of the steps, {\it i.e.}, $x_{\pm i_{cs}}=\pm \alpha(1+2k)/2$  with $i,k=1,\ldots,n$ and $i,k\in\Z^+$. The commutation surfaces and equilibrium points introduced by the function are located when the corresponding domain $\mathcal{D}_i$ is being visited by the trajectory. It is important to mention that if $\alpha$ corresponds to the distance between two continuous equilibrium points given a value of $c$, then, each equilibrium point is located exactly at the middle of two consecutive commutation surfaces. Besides, the equilibria of the system and the domains are located from $(-\infty,\infty)$ due to the fact that the function given by \eqref{ec_round} is not bounded as the commutation law in \eqref{ec_rule2} or \eqref{ec_rule6} are. This can be understand in the following way, if the system increases the size of  its scroll, eventually it crosses to the next or previous $\mathcal{D}_i$ changing the value of  $c$ and  increasing the number of scrolls in the system. This number of scrolls continues increasing according with $t\to \infty$.

In Figure \ref{fig:atractor_round_1} a projection of the trajectory of the system \eqref{ec_sistemABu} given by eqs. \eqref{ec_AB_R3}  \eqref{ec_A} and \eqref{ec_round} into the plane $(x_1,x_2)$ is displayed, presenting each commutation surface and equilibrium point between $-3<x_1<5$ generated from the signal \eqref{ec_round} considering $c=6.3$,  $\alpha=0.6$ and $\mathbf{X}_0=(-0.1,0,0)^T$. Notice that the oscillating behavior is similar as the one based on the commutation surfaces given by eq.~\eqref{ec_rule6}, but $9$ scrolls are presented in this case for 50,000 iterations using the fourth order Runge Kutta method. If the number of iterations are increased so will the number of scrolls.

The relationship between these two parameters satisfies $c/\alpha=10.5$ which is equal to the entry  $a_{31}$ of the  matrix $\mathbf{A}$, resulting in equilibrium points equally located between the commutation surfaces.

\section{Generalized Multistability via PWL systems}

Now based on the previous method for generating multi-scroll attractors,  the following system is considered:
\begin{equation}\label{ec_multistable}
  \mathbf{A}=\left(
    \begin{array}{ccc}
      0 & 1 & 0 \\
      0 & 0 & 1 \\
      -10.5*\nu&-7.0*\nu& -0.7*\nu
    \end{array}
  \right),
\mathbf{B}=\left(
    \begin{array}{c}
      0 \\
      0 \\
      \nu*b_3
    \end{array}
  \right),
\end{equation}

\noindent where $\nu\in\R^+$ is a constant parameter.

With this vector $\bf{B}$ the displacement of the equilibria is also along the $x_1$ axis,  where  $b_3$ commutes according to the $round$ function given by eq.~\eqref{ec_round}.

  \begin{figure}[!t]
  \centering
                    \includegraphics[width=0.45\textwidth]{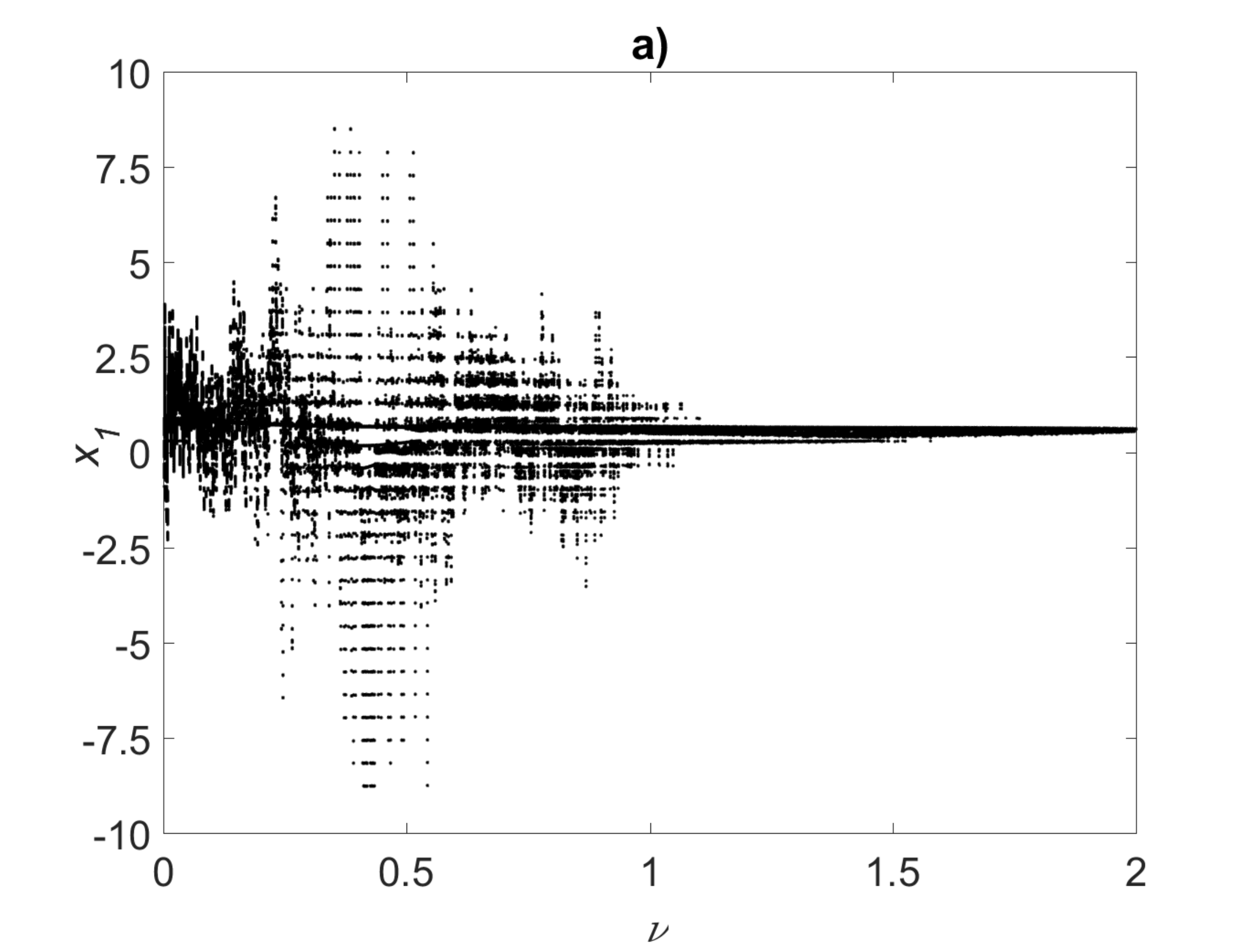}
                    \includegraphics[width=0.45\textwidth]{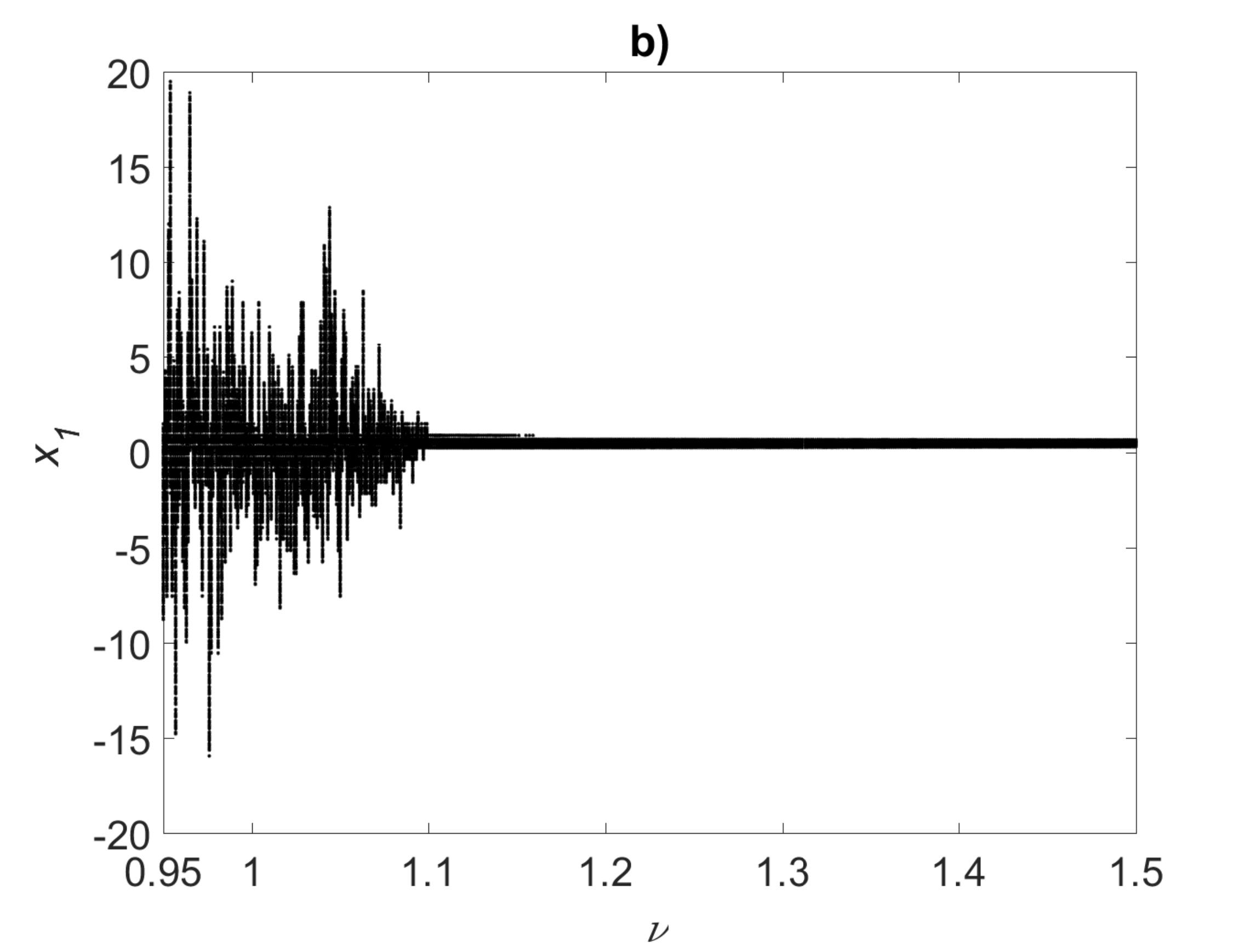}\\
                    \includegraphics[width=0.45\textwidth]{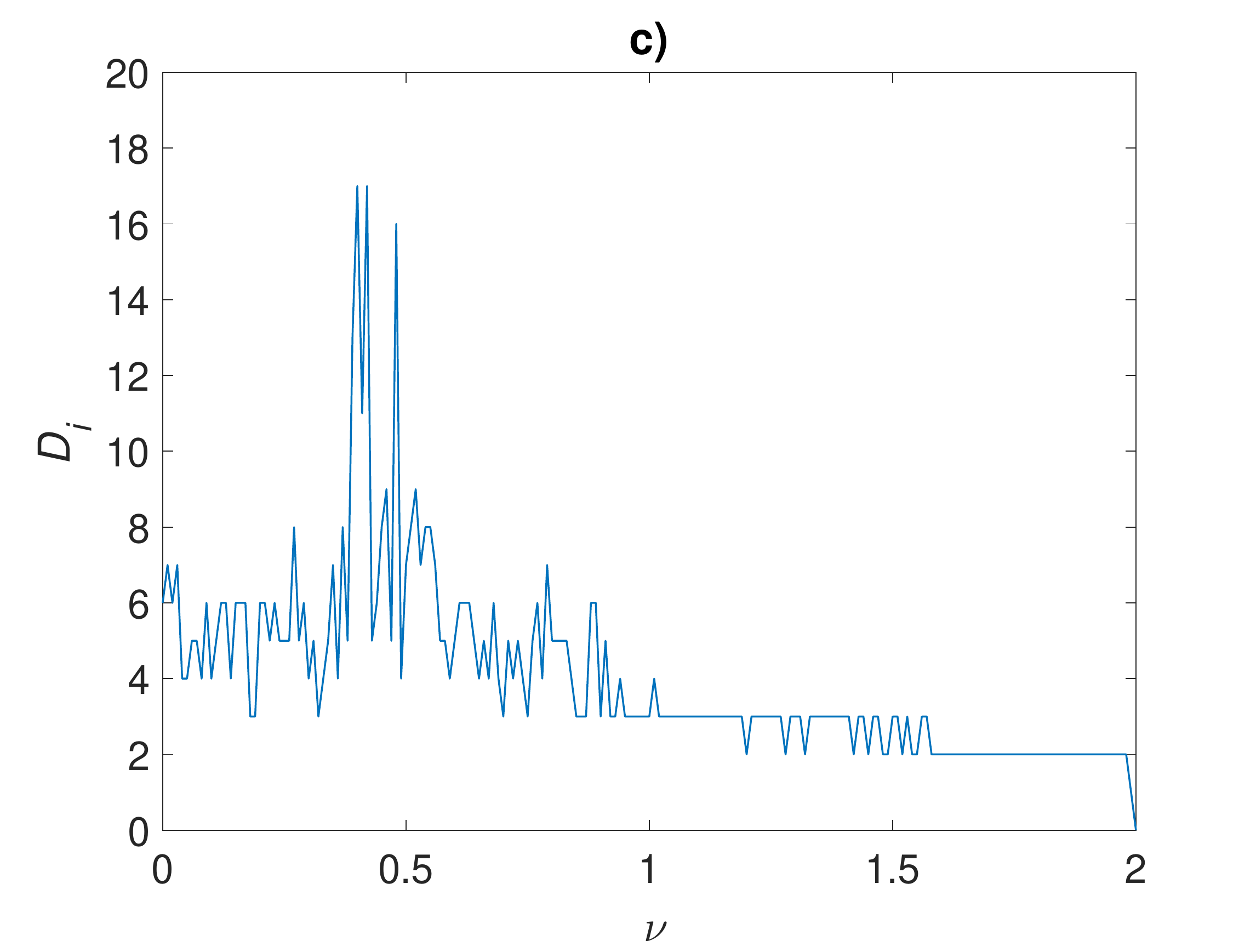}
                    \includegraphics[width=0.45\textwidth]{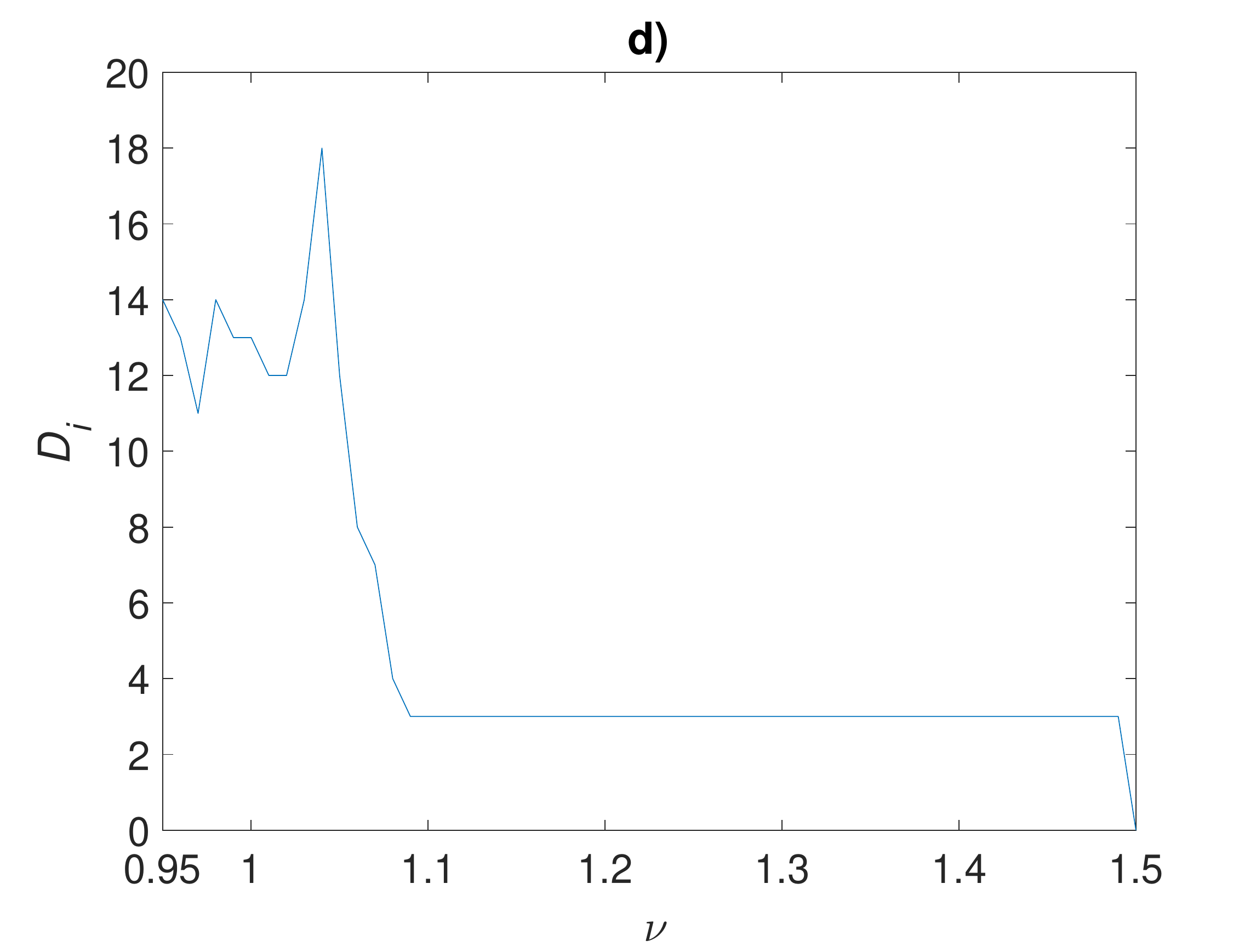}\\

                   \caption{\label{fig:bif} a) Bifurcation diagram of the system given by eqs. \eqref{ec_sistemABu} with \eqref{ec_round} and \eqref{ec_multistable} for the value of $0\leq \nu\leq 2$ for Figure b) depicts the bifurcation for the range $0.95\leq \nu\leq 1.5$ for $1,000,000$ iterations. Figures c) and d) show the number of domains $\mathcal{D}_i$ visited by the trajectory of the systems for the same values of the bifurcation parameters above. The initial condition considered for both diagrams is $\mathbf{X}_0 = (0.7,0,0)$.}
    \end{figure}

\begin{figure}[!t]
  \centering
\includegraphics[width=11cm]{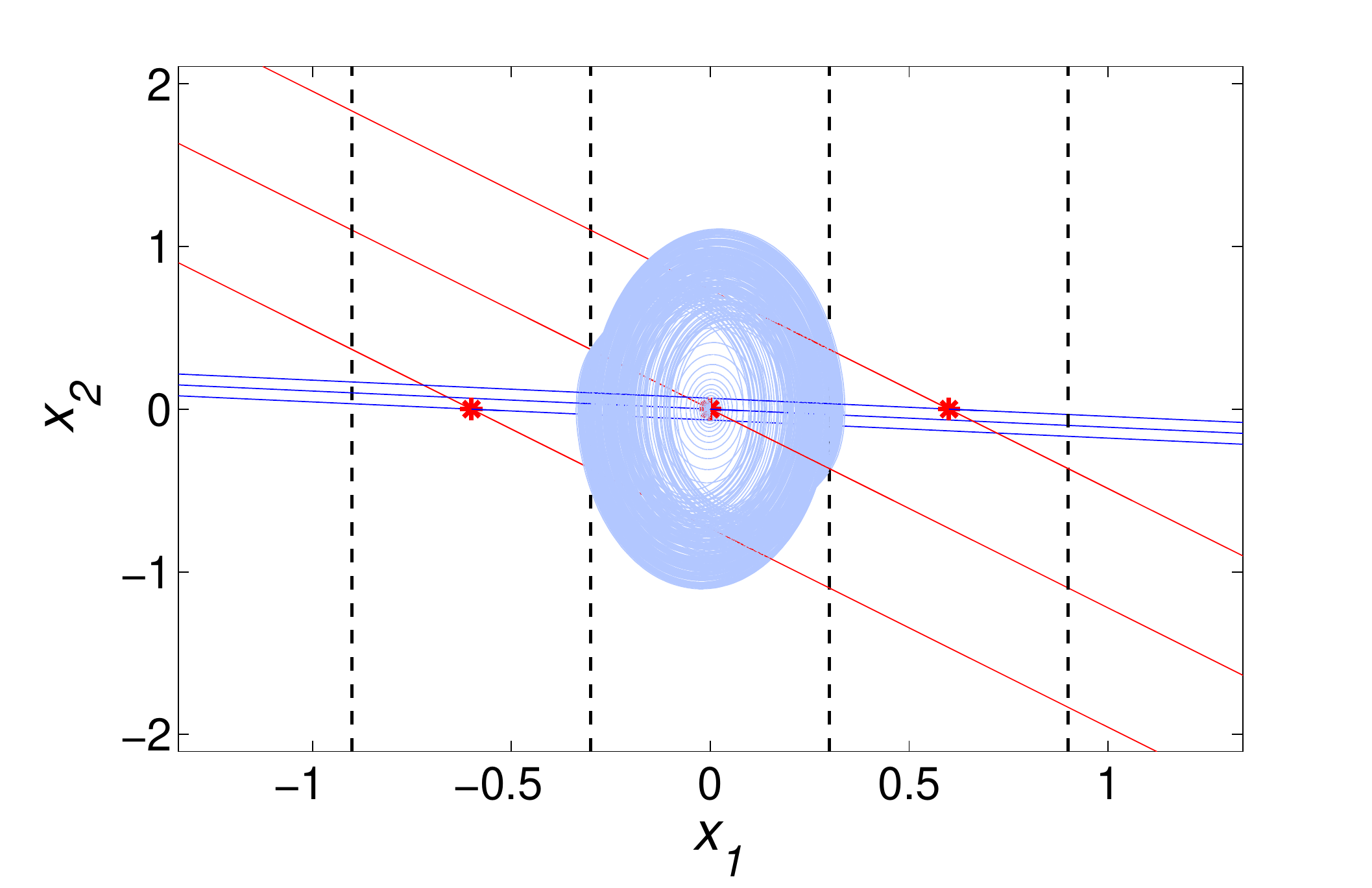}
\caption{\label{fig:multistability_1} Projection of the trajectory of the system  \eqref{ec_multistable} with \eqref{ec_round} onto the $(x_1,x_2)$ plane with $c=6.3$, $\alpha=0.6$ and $\nu=1.42$. Marked with red asterisk the equilibria of the system, and with gray line the commutation surfaces generated by the function  \eqref{ec_round}. The initial condition  of the system is $\mathbf{X}_0=(-0.1,0,0)^T$.}
\end{figure}

By using this approach, the equilibrium  points and the position of the commutation surfaces are preserved. The parameter $\nu$ is used to change the eigenspectra of the linear part of systems, mainly the directions of the stable and unstable manifolds,  so $\nu$ can be taken as a bifurcation parameter. Figures \ref{fig:bif} a) and b)  show the bifurcation diagram of the maximum local at every $\mathcal{D}_i$ of each scroll depicted at $x_1$ for the range of the parameter $0\leq \nu \leq 2$ and a zoomed area at $0.95\leq \nu \leq 1.5$. Both diagrams were calculated by the same initial condition $\mathbf{X}_0 = (0.7,0,0)$, the difference is that Figure \ref{fig:bif} a) was calculated for $10,000$ iterations while Figure \ref{fig:bif} b) for $1,000,000$ iterations.  Additionally Figures \ref{fig:bif} c) and d) show the number of domains $\mathcal{D}_i$ that the system visited for the same values of $\nu$, respectively. Notice that, if the range of $\nu\geq 1.1$ approximately,  the number of domains $\mathcal{D}$ visited remain at the constant value of 3, because the system presents only one attractor located in the inside domain as depicted in Figure \ref{fig:multistability_1}. This can be better appreciated in the zoomed area in Figure \ref{fig:bif} d).

With these parameter values the equilibria of the system is given by $\mathbf{X}^{*}_{0}=(b_3/10.5, 0,0)^T$. Therefore $c=6.3$ and $\alpha=c/10.5$ are assigned to the function in eq.~\eqref{ec_round}.
The eigenspectrum of the matrix $\bf{A}$ depends now on the bifurcation parameter $\nu$, which for the value of $\nu=1.42$ is given by:

\begin{equation}\label{}
\begin{array}{l}
    \lambda(\nu)=\{-1.4164 , 0.2082 \pm i 3.2475\},\\
    \vartheta(\nu)=\{\vartheta_{1,2,3}(\nu)\} = \left\{\left(\begin{array}{c}0.3771 \\  -0.5342\\0.7566\end{array}\right)
\left(\begin{array}{c}0.0892 \pm i 0.0115\\ 0.0187 \pm i 0.2919\\-0.9520\end{array}\right)\right\},
\end{array}%
\end{equation}%

\noindent proving that the Definition \ref{def_USD3} is satisfied. Considering these values, the system results in an interesting multistable state phenomena due to the round function and the direction of their eigenvectors as depicted in Figure \ref{fig:multistability_1} for the initial condition $\mathbf{X}_0=(-0.1,0,0)^T$. The  attractor  is oscillating near the equilibria in the origin marked in red asterisk due to the initial condition given. However there is no oscillation near the adjacent equilibrium points, the reason of this resides on the eigenvectors, as they are not located in the same way as in the previous examples, {\it i.e.}, the stable manifold of the oscillating domain, doesn't match and cross with the stable manifolds of the adjacent domains.

\begin{figure}[!t]
                    \includegraphics[width=0.45\textwidth]{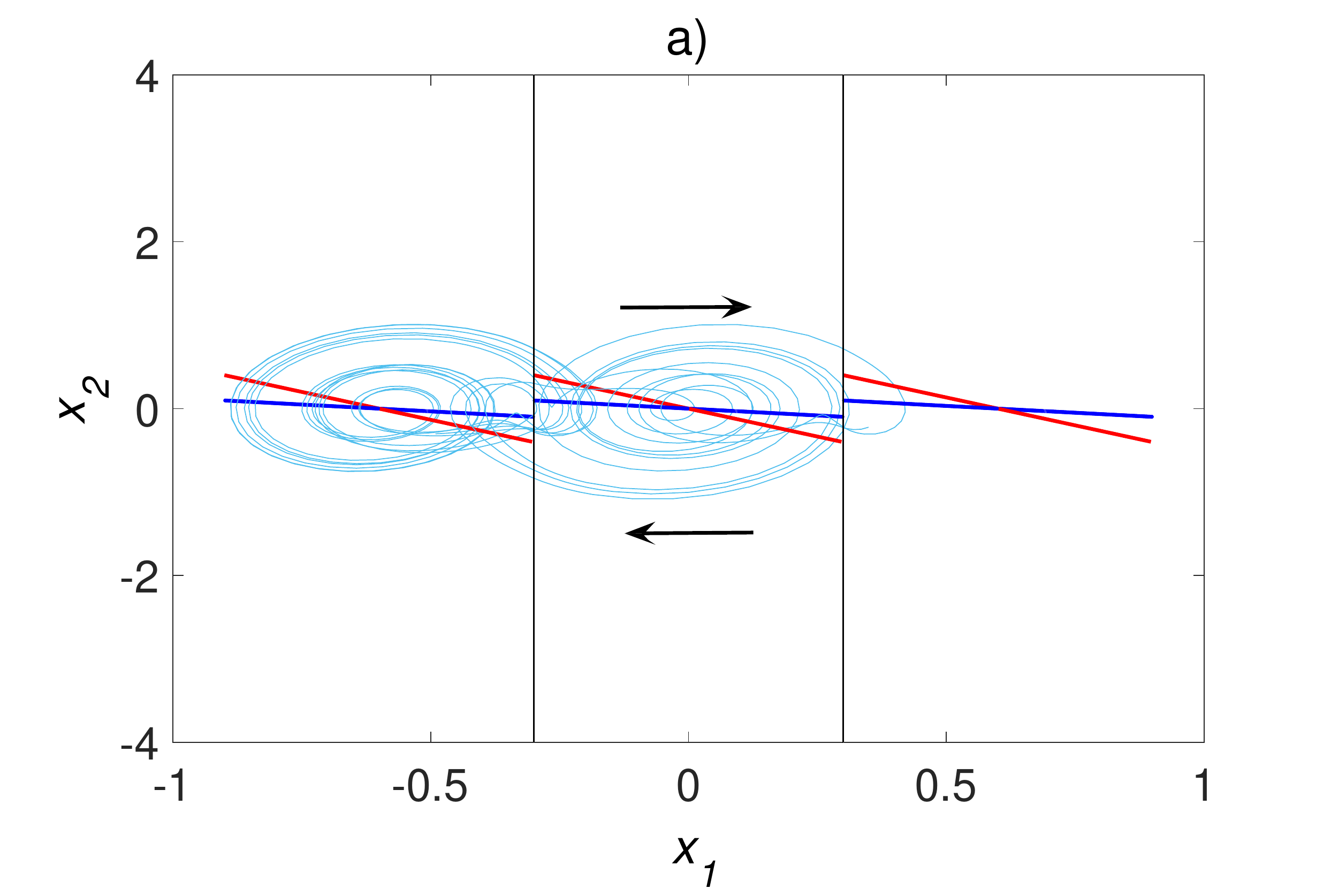}
                    \includegraphics[width=0.45\textwidth]{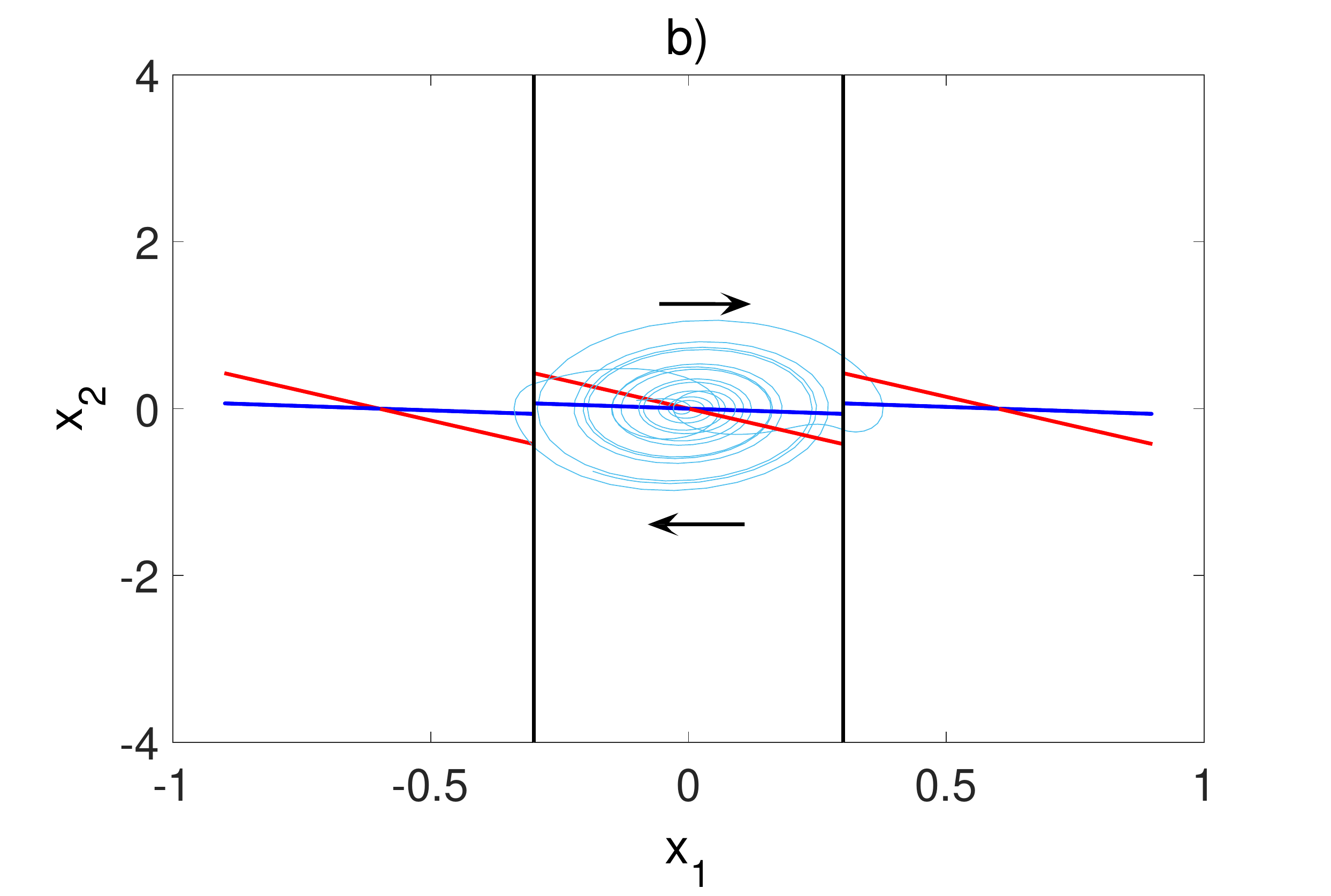}\\
                    \includegraphics[width=0.45\textwidth]{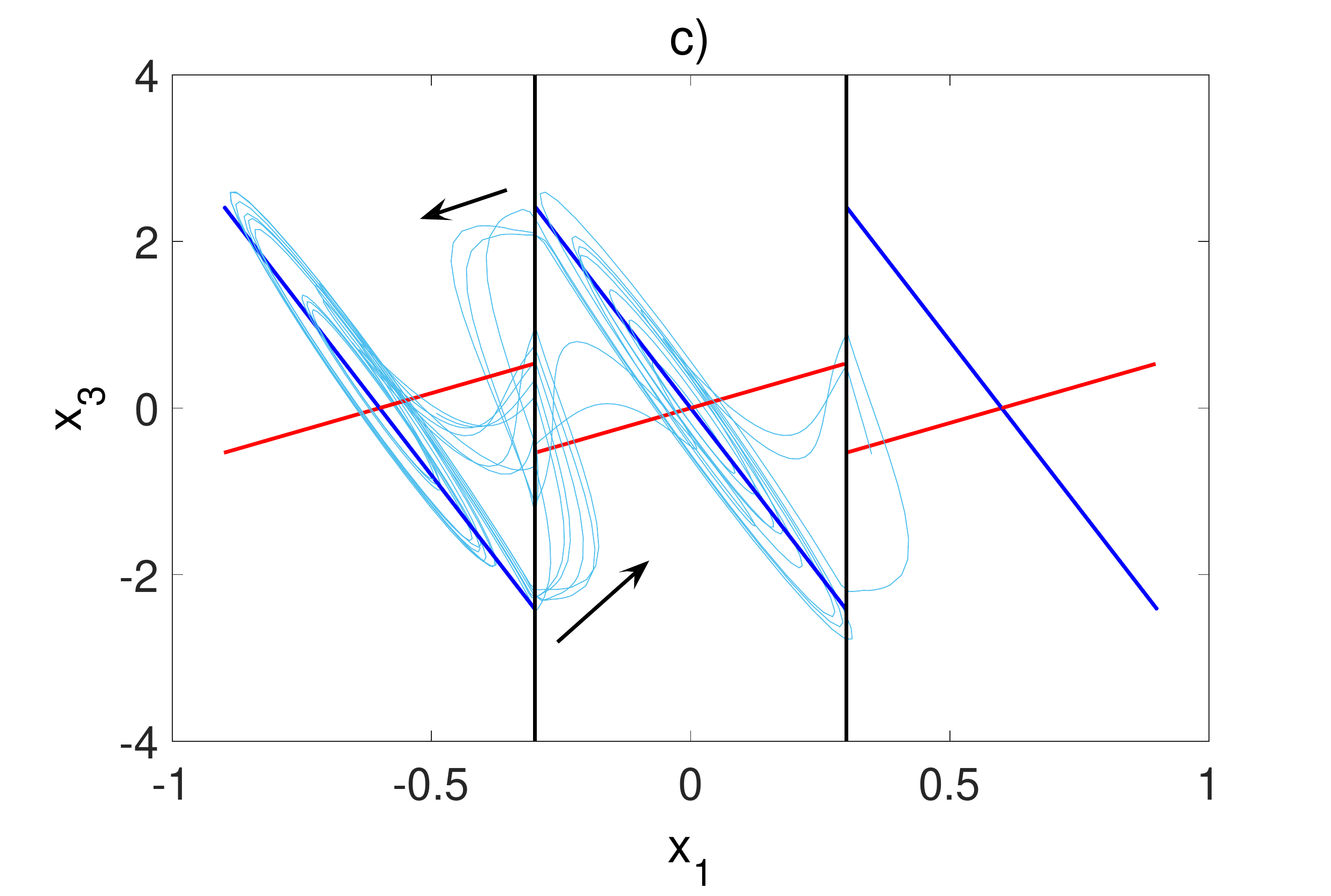}
                    \includegraphics[width=0.45\textwidth]{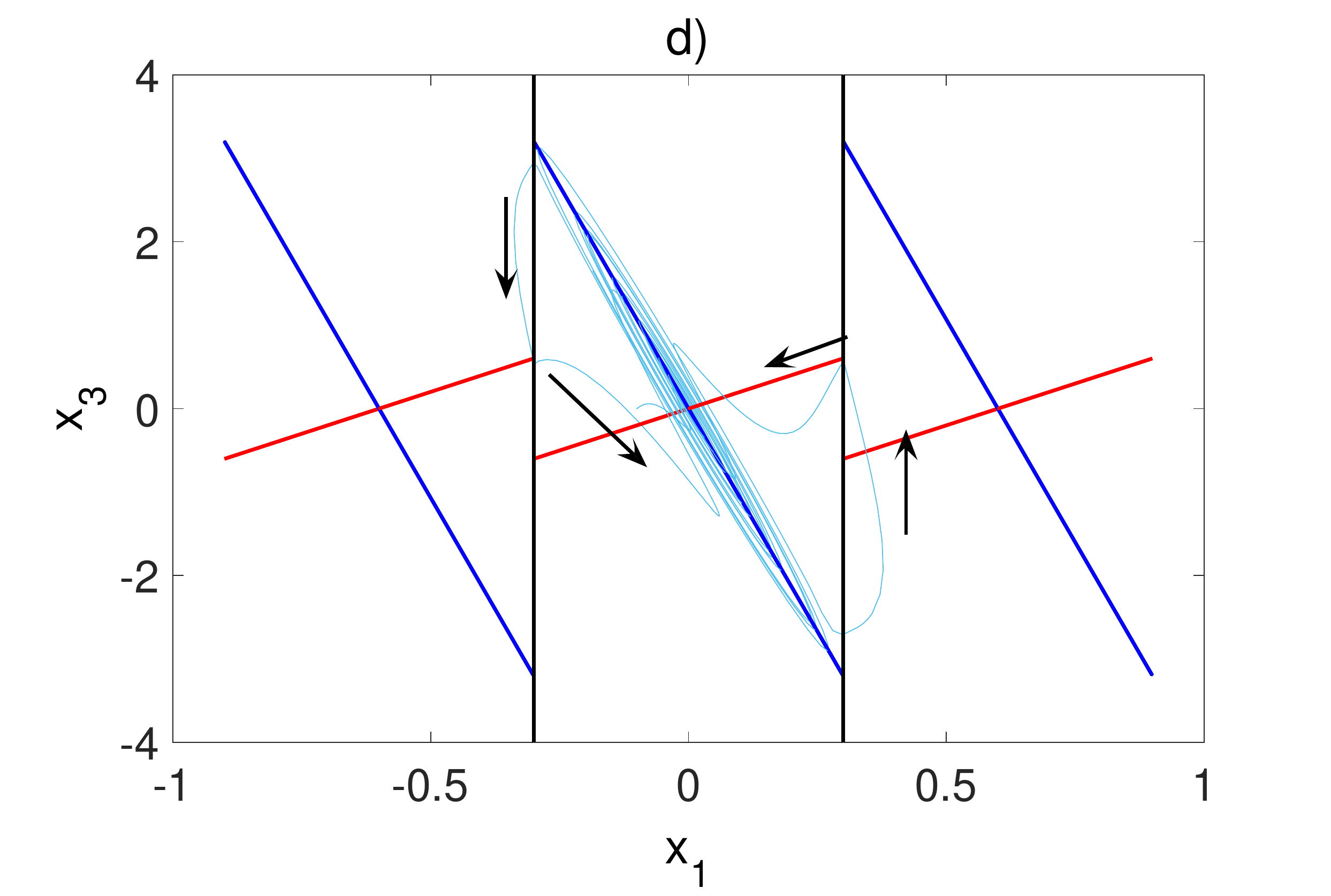}
                   \caption{\label{fig:variedades} Trajectory projections for the system given by eq. \eqref{ec_multistable}, with $c=6.3$ and $\alpha=0.6$. For $\nu=1$  a) onto the plane $(x_1,x_2)$, c) onto the plane $(x_1,x_3)$. For $\nu=1.42$  b) onto the plane $(x_1,x_2)$, d) onto the plane $(x_1,x_3)$. Marked with  green line the commutation surfaces generated by the function  \eqref{ec_round}, with red line the complex eigenvector and with black  line the real eigenvector. The black arrows show the  trajectory direction. }
    \end{figure}

This can be easily observed in Figure \ref{fig:variedades} as it is next explained. Figures \ref{fig:variedades} a) and b) present a projection of a trajectory onto the plane $(x_1,x_2)$ for the systems  \eqref{ec_sistemABu} with \eqref{ec_round} and \eqref{ec_multistable}, both initialized with the same initial condition near the origin $\mathbf{X}_0=(-0.1,0,0)^T$, but considering the values in the bifurcation parameter of  $\nu=1$ and $\nu=1.42$, respectively. Notice in the projections that as the time increases the oscillating clockwise trajectory on both systems attractor grows larger, until eventually the trajectories on the scrolls cross the commutation surface close to  the unstable manifold of $E^u$ marked in blue lines. Apparently both systems trajectory cross the commutation surface plane $x_{1_{cs}}$ marked with black lines in a neighbourhood near to the intersection of the unstable and stable manifolds marked with blue and  red lines (depending if the trajectory is escaping or entering the domain), respectively. However, by observing different projections of the attractors, for example the projection onto the plane $(x_1,x_3)$ in Figures \ref{fig:variedades} c) and d), it can be appreciated that the direction of the manifold has been slightly changed due to the variation of the parameter from $\nu=1$ to $\nu=1.42$.
Take a closer look at the graph in Figure \ref{fig:variedades} c) between $-1<x_1< -0.3$. In this domain $\mathcal{D}_{-1}$ the trajectory of the system is entering from the upper part close to $x_3 \approx 2$ where the black arrow depicts the direction of the crossing in the intersection of the unstable manifold with the commutation surface ( $E^u\cap x_{-1_{cs}}$). After entering to $\mathcal{D}_{-1}$ the trajectory is directed below the stable manifold $E^s$ in this domain marked with the red line, and then crosses to the domain $\mathcal{D}_{0}$ near the intersection of the next stable manifold in $\mathcal{D}_{0}$ in $x_3\approx -0.9$. Notice that some of the trajectories entering this domain end up oscillating into the scroll as the arrow in the lower part depicts (this phenomena can also be seen in the projection of the multi-scroll attractor in Figures \ref{fig:Bi} b) and d)). Nevertheless, some of the trajectories instead of reaching the scrolling plane in $\mathcal{D}_{0}$ are redirected again to $\mathcal{D}_{-1}$ and oscillate in the scroll in the unstable manifold $E^u$.

This behaviour is completely different in the projection of the attractor in Figure \ref{fig:variedades} d) for $\nu = 1.42$. Here, when the trajectory escapes the domain $\mathcal{D}_{0}$ near $x_3 \approx 3$ and enters $\mathcal{D}_{-1}$, the trajectory is directed towards the location of the stable manifold and crosses near the intersection $E^s\cap x_{-1_{cs}}$ at approximately $x_3\approx0.1$. After crossing the trajectory is redirected towards the scroll in $\mathcal{D}_{0}$ as the arrows depicts the direction. This process repeats continuously from the three consecutive domains in which the trajectory of the system lies (Notice this also from Figure \ref{fig:bif} b)).

\subsection{Transition from the multi-scroll to multistable state phenomenon}

Now, in order to understand the phenomenon and visualize the exact location of the  intersection of $E^s$ and  $E^u$  along with the points belonging to { the attractor $\mathfrak{A}$ at the commutation surface} $x_{1_{cs}}$,
a Poincar\'e plane was implemented exactly at the commutation surface. First, the Poincar\'e  plane is defined as $\Sigma := \{ (x_1,x_2,x_3)\in \R^3: \mu_1x_1 +\mu_2x_2+\mu_3x_3+\mu_4=0 \}$, where $\mu_1,\ldots,\mu_4\in\R$ are the coefficients of an hyperplane equation whose values are considered depending on the location under study, which in this case it will be in the commutation surfaces $x_{i_{cs}}$ with $i\in \Z $ guaranteeing
$\mathfrak{A}\cap \Sigma \neq \emptyset$. The crossing events of interest are $\{ \phi^{t_1}_{in}(\mathbf{X_0}), \phi^{t_2}_{out}(\mathbf{X_0}),\phi^{t_3}_{in}(\mathbf{X_0}), \phi^{t_4}_{out}(\mathbf{X_0}), \ldots, \phi^{t_{m-1}}_{in}(\mathbf{X_0}), \phi^{t_m}_{out}(\mathbf{X_0})\}\in \Sigma$ with $m \in \Z^+$. Where  $m$ corresponds to the total of crossing events in $\Sigma$, and $\phi^j_f$ correspond to the $j$-th intersection of $\mathfrak{A}\cap\Sigma$ in the $f = in,out$ direction. { The sub-index $out$ corresponds to trajectories  that cross $\Sigma$ with $dx_1/dt>0$, and $in$ corresponds to trajectories  that cross $\Sigma$ with $dx_1/dt<0$.}

Figures \ref{fig:poincare_sections} a) and b) present the projections of $\Sigma := \{(x_1,x_2,x_3)\in \R^3: x_1-x_{1_{cs}}=0\}$ for the values of $\nu = 1$ and $\nu = 1.42$, respectively. Three points can be noticed from these projections.

\subsubsection*{I.- Distribution of the intersections with the Poincar\'e plane}
\begin{figure}[!t]
  \centering
    \includegraphics[width=9cm]{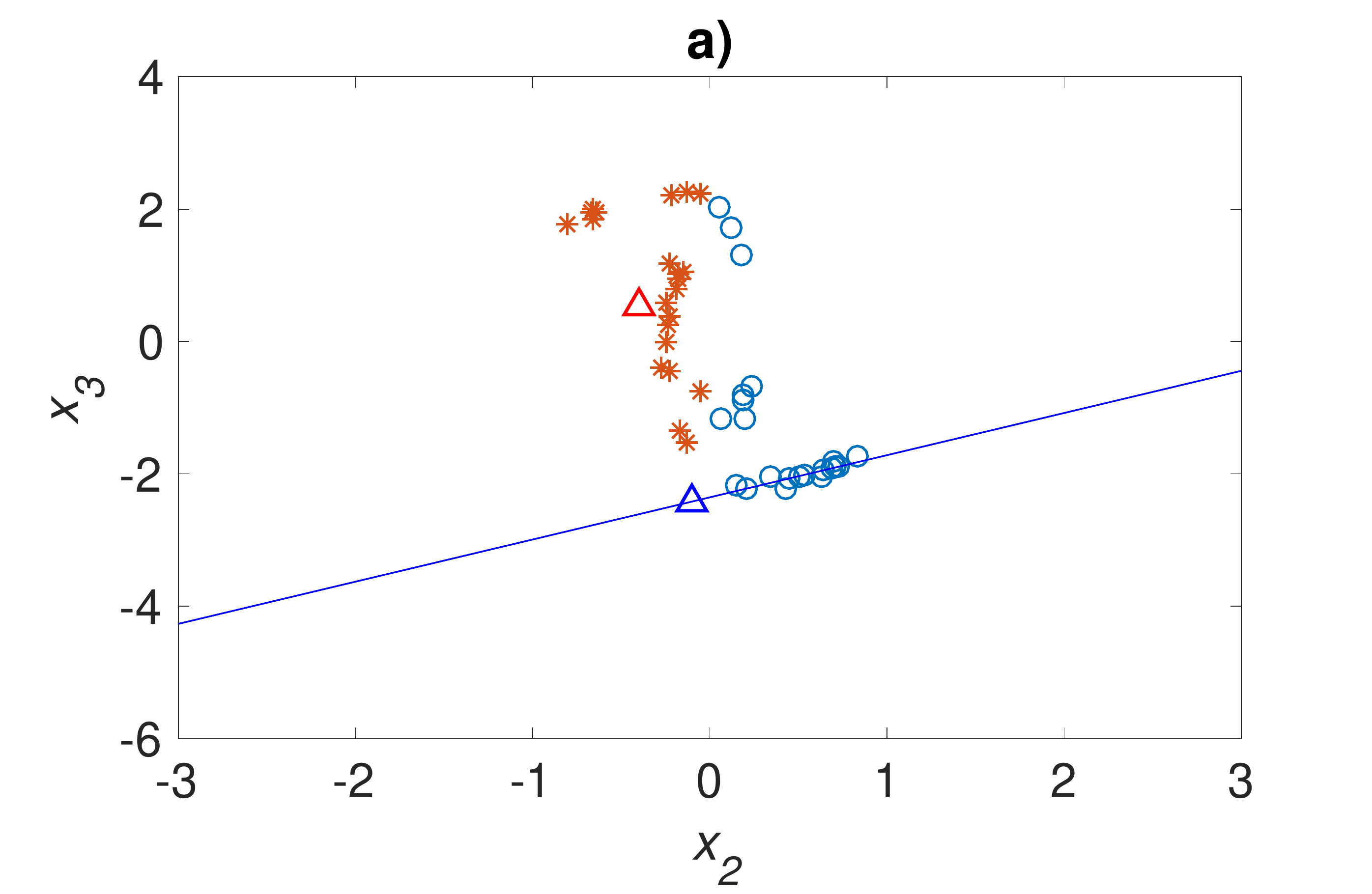}
    \includegraphics[width=9cm]{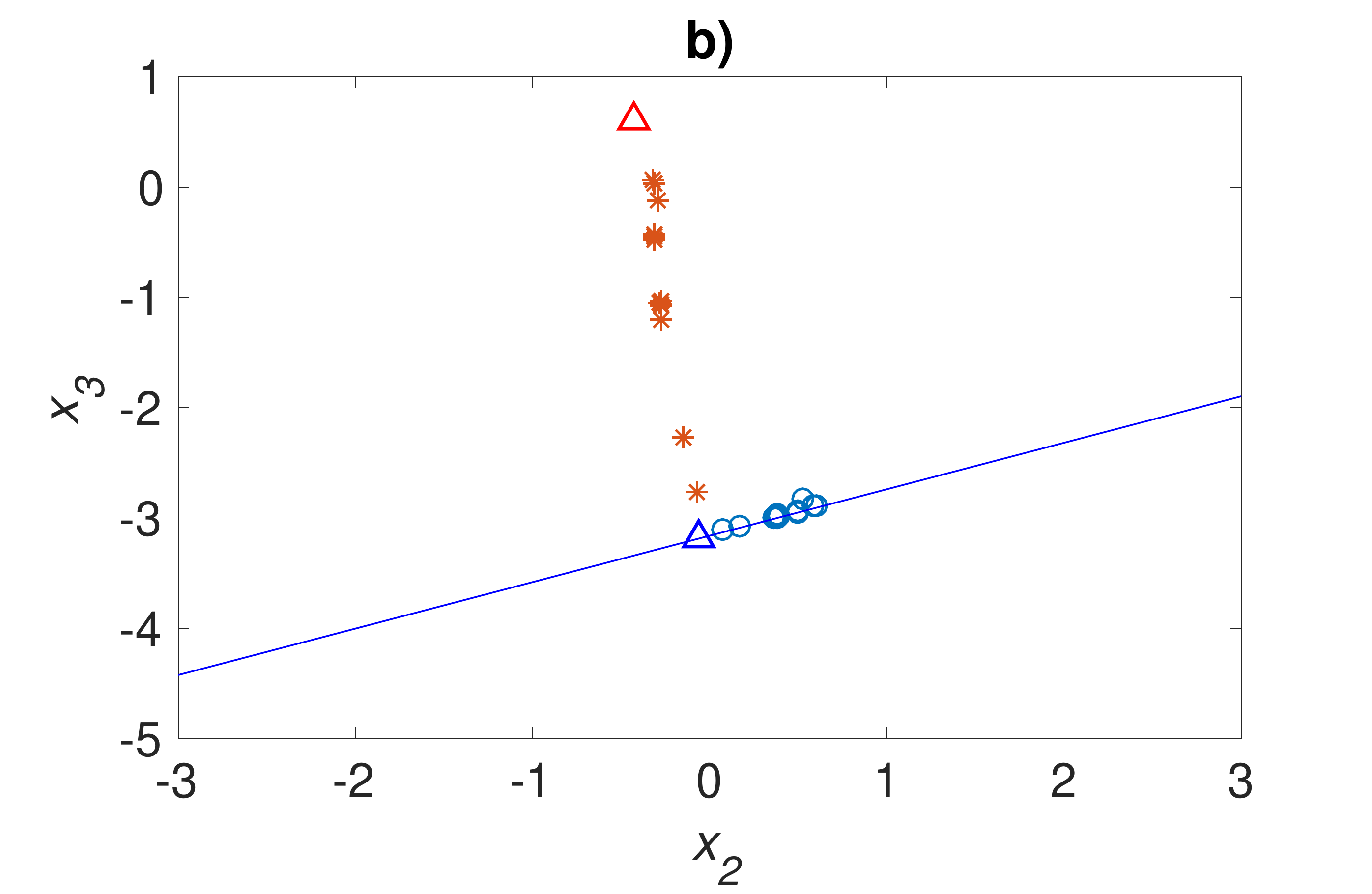}
\caption{\label{fig:poincare_sections}  Intersections of the trajectory of the  system  \eqref{ec_sistemABu} with \eqref{ec_round} and \eqref{ec_multistable} with the commutation surface $x_{1_{cs}}$ for a)  $\nu=1$, b) $\nu=1.42$. Marked with blue circles the trajectories exiting $\mathcal{D}_1$ near $E^u\cap x_{1_{cs}}$. The orange asterisk represent the trajectories entering $\mathcal{D}_1$. The red triangle stands for $E^s\cap x_{1_{cs}}$, and the blue line corresponds to the intersection of unstable manifold and commutation surface.}
\end{figure}

When the trajectory of $\mathfrak{A}$ is increasing its scroll size it exits the current domain $\mathcal{D}_i$ to $\mathcal{D}_{i+1}$ (or  $\mathcal{D}_i$ to $\mathcal{D}_{i-1}$) through  the commutation surface near the intersection of the unstable manifold with the Poincar\'e plane ($E^u\cap\Sigma$), as it can be appreciated with the  intersection points of the trajectory $\phi^{t_j}_{out}(\mathbf{X}_0)$ marked in blue circles. The blue triangle corresponds to the intersection of $Real(\vartheta_2(\nu)) \cap \Sigma$.
The blue line appearing  in both Figures \ref{fig:poincare_sections} a) and b), corresponds to the intersection of the unstable manifold with the Poincar\'e plane  $E^u=Span\{Real(\vartheta_2(\nu)), Imag(\vartheta_2(\nu))\} \cap \Sigma$. Notice that most of the trajectories are crossing near this section due to the scrolling behaviour in or near the unstable manifold. Nevertheless, Figure \ref{fig:poincare_sections} a) presents a region of escaping intersection points $\phi^{t_j}_{out}(\mathbf{X}_0)$ not near the unstable manifold, which comes out as one of the main difference between the escaping points in Figure \ref{fig:poincare_sections} b) for a larger value of the bifurcation parameter. Also notice that the entering intersection points $\phi^{t_j}_{in}(\mathbf{X_0})$,  marked with orange asterisks for $\nu = 1$ are located around the intersection of the stable manifold $E^s\cap \Sigma$ marked with the red triangle in Figure \ref{fig:poincare_sections} a), but in Figure \ref{fig:poincare_sections} b), the events are located below this intersection  in an apparently ranked way.

\subsubsection*{II.- Distance of the intersections with the Poincar\'e plane and the intersection of the manifolds}

\begin{figure}
                    \includegraphics[width=0.45\textwidth]{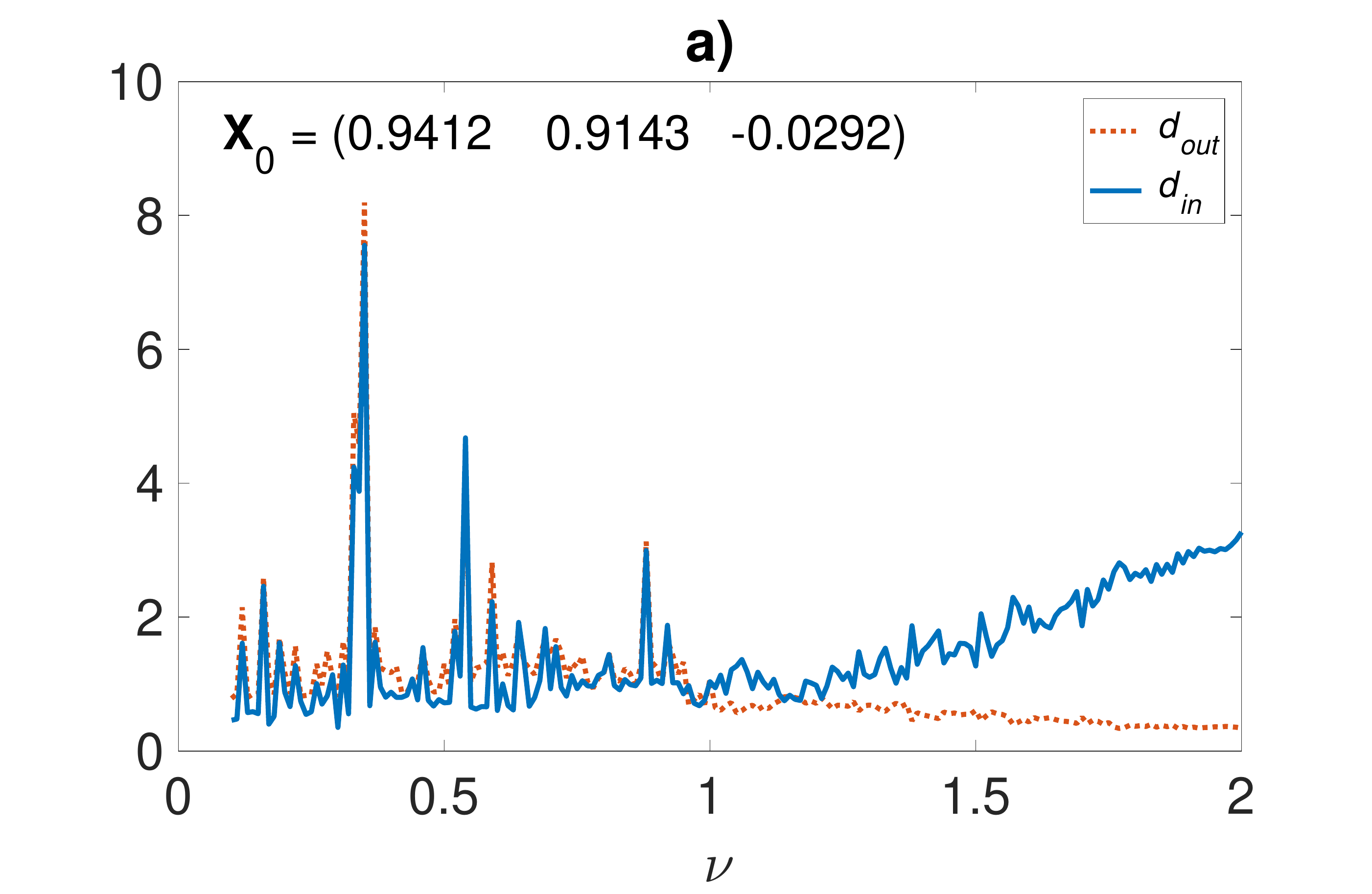}
                    \includegraphics[width=0.45\textwidth]{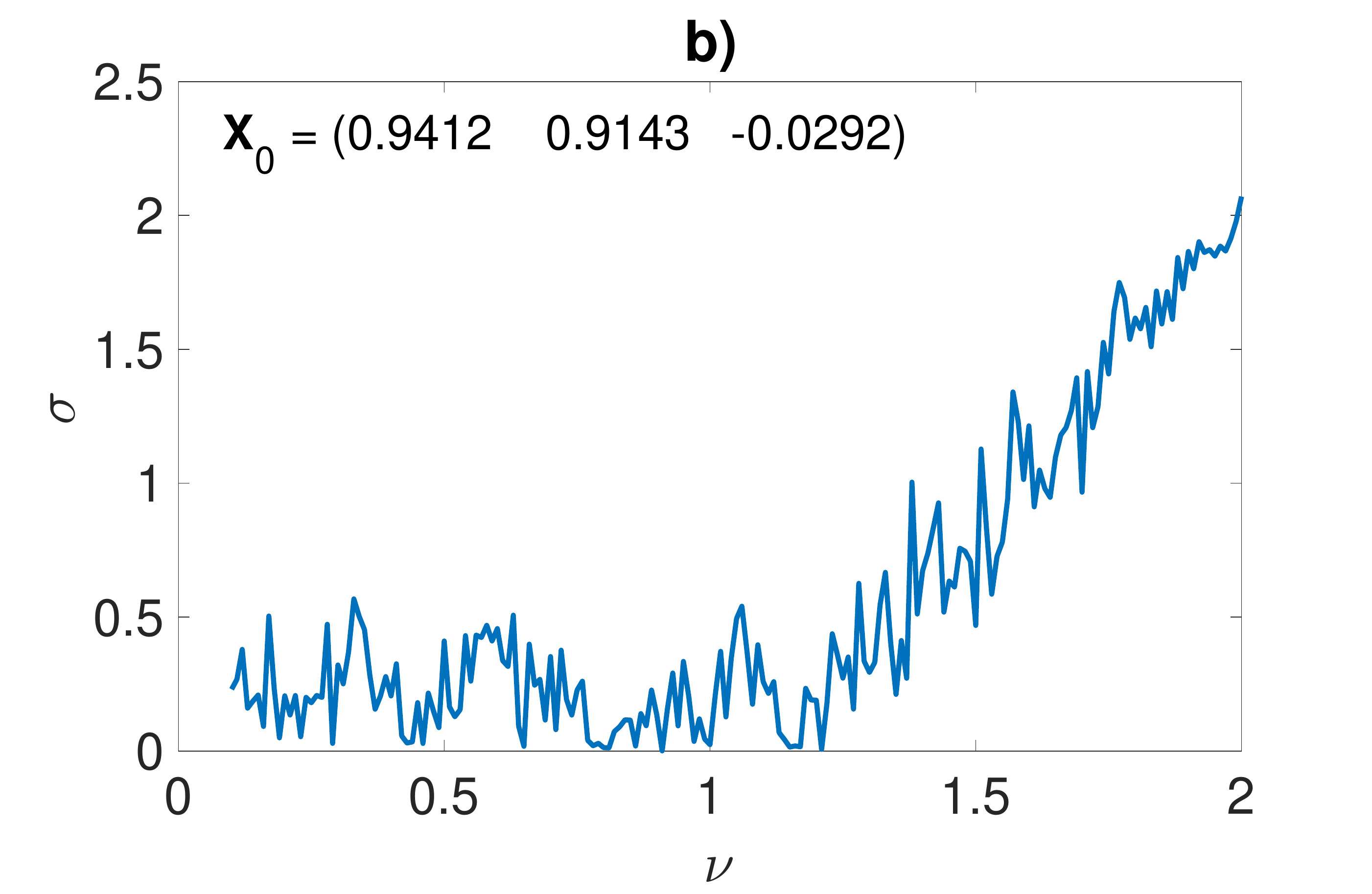}\\
                    \includegraphics[width=0.45\textwidth]{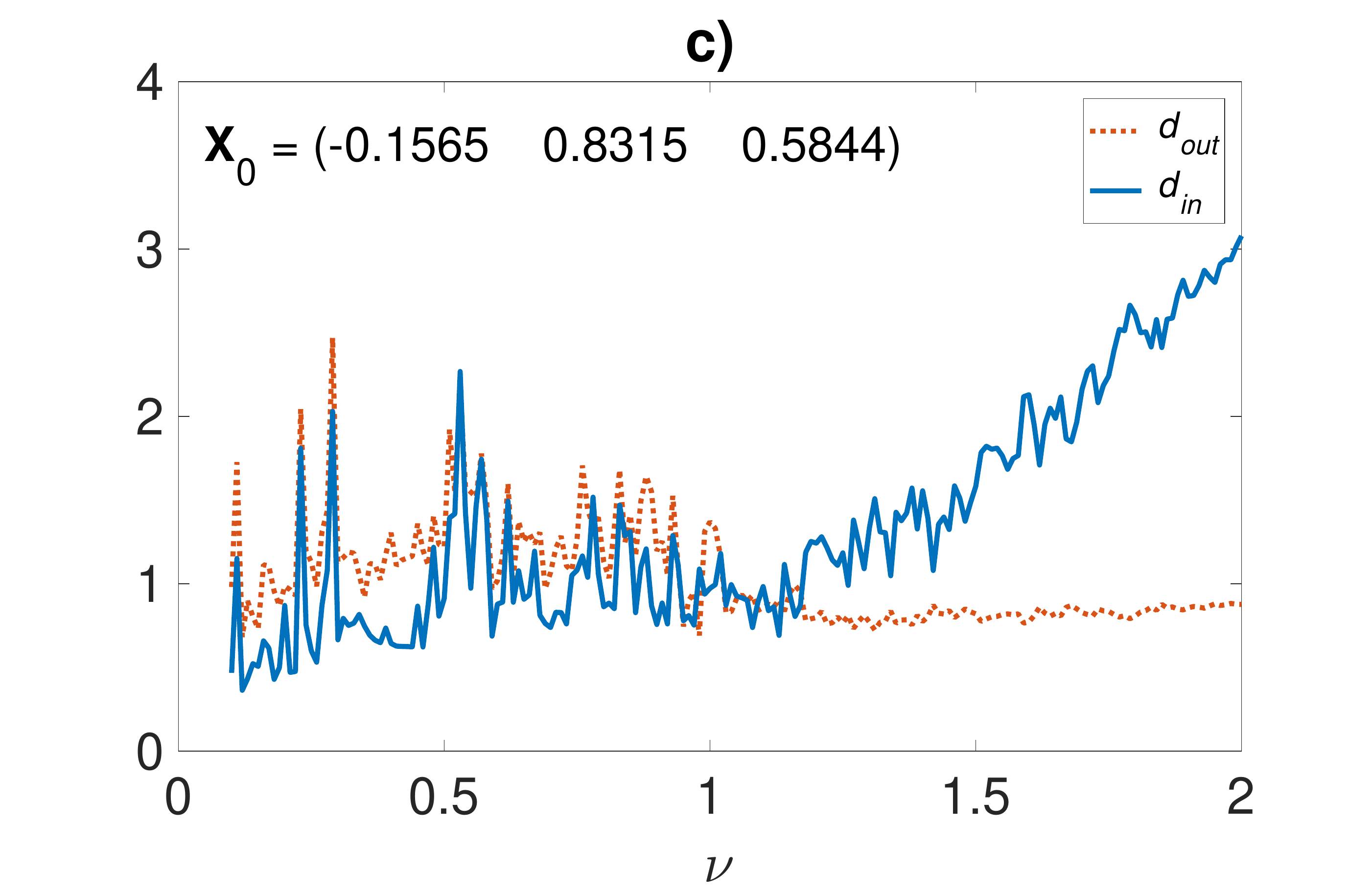}
                    \includegraphics[width=0.45\textwidth]{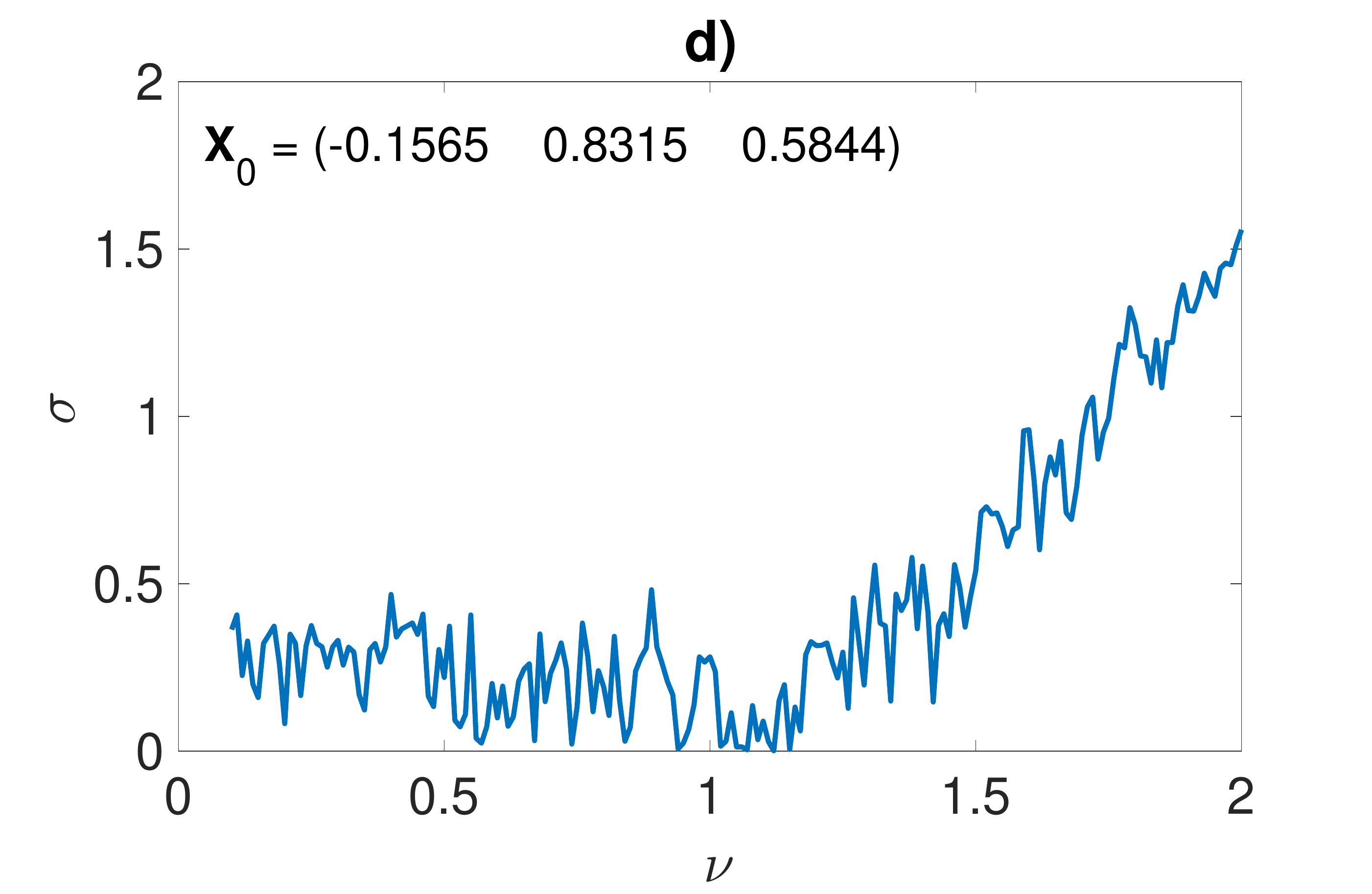}
                   \caption{\label{fig:distancia} Distance calculated from eq. \eqref{distancia_cruces} between the crossing trajectory events and the crossing of the corresponding manifold whether the trajectory is displacing inwards $d_{in}$ (continuous blue line) or outward $d_{out}$ (dotted orange line). The initial conditions considered are: a) $\mathbf{X}_0 = (-0.9412,0.9143,-0.0292)$, c) $\mathbf{X}_0 = (-0.1565,0.8315,0.5844)$. Figures b) and d) depict the standard deviation $\sigma$ from eq. \eqref{STD} between the  distances for the each corresponding set of initial conditions.}
    \end{figure}

To determine the relationship among the bifurcation parameter and the multi-scroll or multistable solutions, the distance between these crossing trajectory events and the crossing of the corresponding manifold (whether the trajectory is displacing in or out) was calculated as follows:

\begin{equation}\label{distancia_cruces}
\begin{array}{l}
    d_f = d(\phi^{t_j}_{f}(\mathbf{X_0}),E^k\cap\Sigma) =
    \sqrt{(x_1^\phi-x_1^E)^2+(x_2^\phi-x_2^E)^2+(x_3^\phi-x_3^E)^2},
\end{array}%
\end{equation}%

\noindent where $j \in \Z+$ corresponds to the $j$-th crossing event of the trajectory in the $f=in,out$ direction for the $k=s,u$ manifold intersection. The triplet $(x_1^\phi,x_2^\phi,x_3^\phi)$ represents the coordinate of the $j$-th corresponding crossing event of the trajectory, and $(x_1^E,x_2^E,x_3^E)$ the coordinate of the intersection $E^k\cap\Sigma$ (which were depicted with blue and red triangles in Figure \ref{fig:poincare_sections}). The results of the distances for $d_{out}$ and $d_{in}$ are depicted in Figure \ref{fig:distancia} in orange dotted line and in blue continuous line, respectively. The range considered in the experiment consists of values of the bifurcation parameter $0.1\leq \nu \leq 2 $ with a spacing of $0.01$ and an initial condition $\mathbf{X}_0 = (-0.9412,0.9143,-0.0292)$ for Figure \ref{fig:distancia} a) and $\mathbf{X}_0 = (-0.1565,0.8315,0.5844)$ for Figure \ref{fig:distancia} c).

Both  distances present a common behaviour for $\nu\leq 1.2$, {\it i.e.}, the distance  $d_{out}$ loses its spiking variational behaviour and becomes smoother, while $d_{in}$ starts to increment in comparison with $d_{out}$. To quantify this variations among the intersections points the standard deviation of the distances is represented in  Figure \ref{fig:distancia} b) and d) for the initial conditions previously mentioned, and it is descried as follows:

\begin{equation}\label{STD}
\begin{array}{l}
\sigma = \sqrt{\frac{1}{N-1}\sum_{i=1}^{N}|d_{in}-\mu_{in}| ^2 +\frac{1}{N-1}\sum_{i=1}^{N}|d_{out}-\mu_{out}|^2},
\end{array}%
\end{equation}%

\noindent  where $N$ corresponds to the number of values of $\nu$ evaluated (for the range discussed above $N = 1901$), and $\mu$ is the mean of $d_{j}$. These values were calculated by the $std(d_{in},d_{out})$ MATLAB function.

For both initials conditions Figure b) and d) present a relation between $d_{out}$ and  $d_{in}$ depicted by the values of $\sigma$  in the range of 0 and $\approx0.5$ for $\nu< 1.2$. However, $\nu\geq 1.2$ the standard deviation among the distances increases. This  correlates with the previous result of the bifurcation diagram in Figure \ref{fig:bif} for the same values of $\nu>1.2$.

\subsubsection*{III.- Eigenspectra and manifold direction due to the bifurcation parameter $\nu$}

The last point to address is that the distance between the stable and unstable manifold crossings in the Poincar\'e plane (marked with red and blue triangles for $E^s\cap\Sigma$ and $Real(\vartheta2)\cap\Sigma$, respectively) is smaller in Figure \ref{fig:poincare_sections} a) than the one in Figure \ref{fig:poincare_sections} b).

\begin{figure}[!t]
  \centering
\includegraphics[width=11cm]{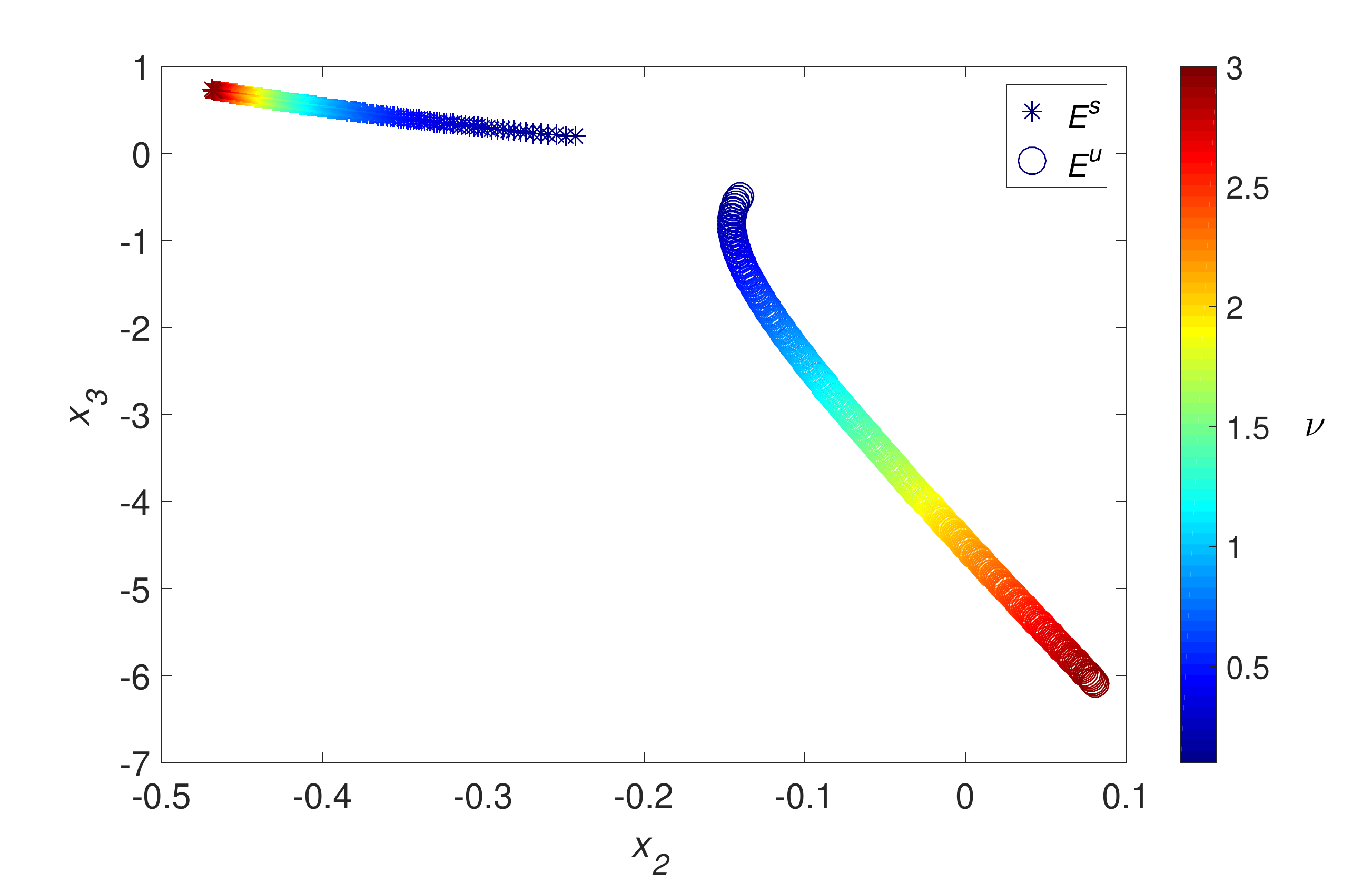}
\caption{\label{fig:poincare_eingevector} Intersections of the stable and unstable manifolds $E^s$ and $Real(E^u)$ with the commutation surface $x_{1_{cs}}$ for  $0.25\leq \nu\leq 3$.}
\end{figure}

To study in more detail this displacement in space of the manifolds, an analysis of the system's eigenspectra due to the variation of the bifurcation parameter was implemented.  The manifolds direction and their crossing position with the commutation surface was analysed with the Poincar\'e plane. The range of the bifurcation parameter considered is of $0.25 \leq \nu \leq 3$, from which for a step of $0.01$ the set of eigenvectors were calculated. Figure \ref{fig:poincare_eingevector} displays the crossing locations due to this variation. And it can be appreciated that the intersection of the manifolds $E^i\cap \Sigma$ with $i = s,u$ (marked with asterisk and circles,respectively) are increasing their separation regarding the variation of $\nu$  shown in the color gradient. This result correlates with the previous analysis of the distance and the escaping locations of the system's trajectory.

\begin{figure}[!t]
  \centering
\includegraphics[width=11cm]{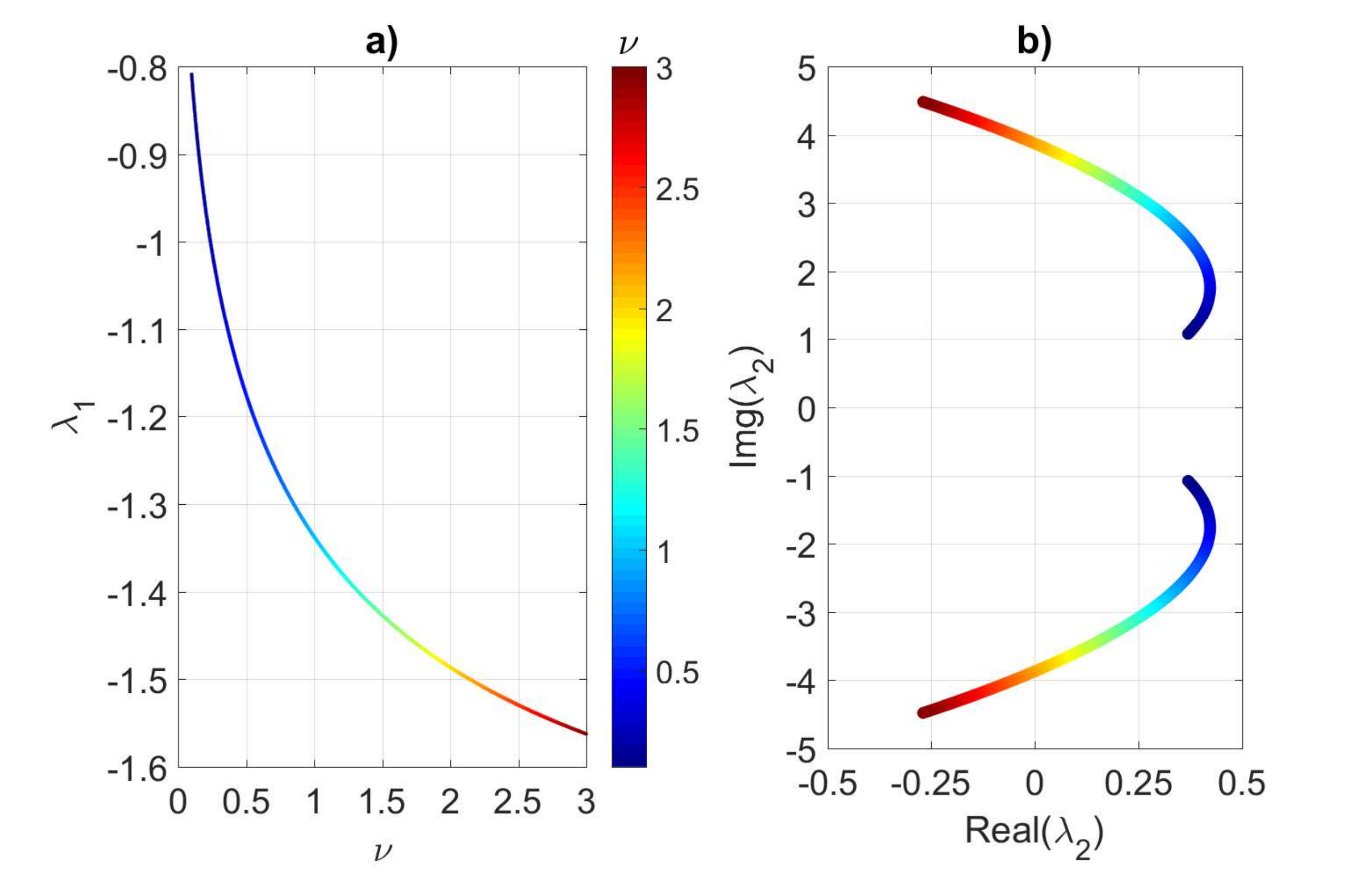}
\caption{\label{fig:eigenvalores} Eigenvalues of the system  \eqref{ec_multistable} with \eqref{ec_round} for  $0.25\leq \nu \leq 3$. Figure a) depicts the real eigenvalue $\lambda_1$, while b) depicts the complex conjugated $\lambda_{2,3}$. }
\end{figure}

In the same context, in Figure \ref{fig:eigenvalores}  are depicted the eigenvalues of the system \eqref{ec_multistable} $\lambda_i$ with $i=1,2,3$ for the same range of $0.25 \leq \nu \leq 3$ resulting in an interesting contrast in their values. The complex conjugate eigenvalues start to decrement  their real part until they become negative for values of $\nu>2.1$ (not satisfying Definition 2.1), and at the same time the imaginary part of the complex conjugate increases (ensuring an augment in the frequency of oscillation). On the other hand,  $\lambda_1$, remains negative with a decreasing magnitude. This results in unstable UDS of the type I equilibria (for $\nu<2.1$), that present a stretch oscillation in the scroll with a stronger attraction towards $E^s$.

Summarizing, the variation of the number of scrolls and domains visited  due to the values in the bifurcation parameter (Figure \ref{fig:bif}), the distance of the intersecting locations of the system's trajectory in the commutation surfaces with respect to the location of the intersection of manifolds (Figure \ref{fig:poincare_sections}), are caused  by the variation of the eigenvalues of the system (Figure \ref{fig:eigenvalores}), resulting in  the  transition between unstable multi-scroll structures  and the multistable structure.

\begin{figure}[!t]
  \centering
\includegraphics[width=11cm]{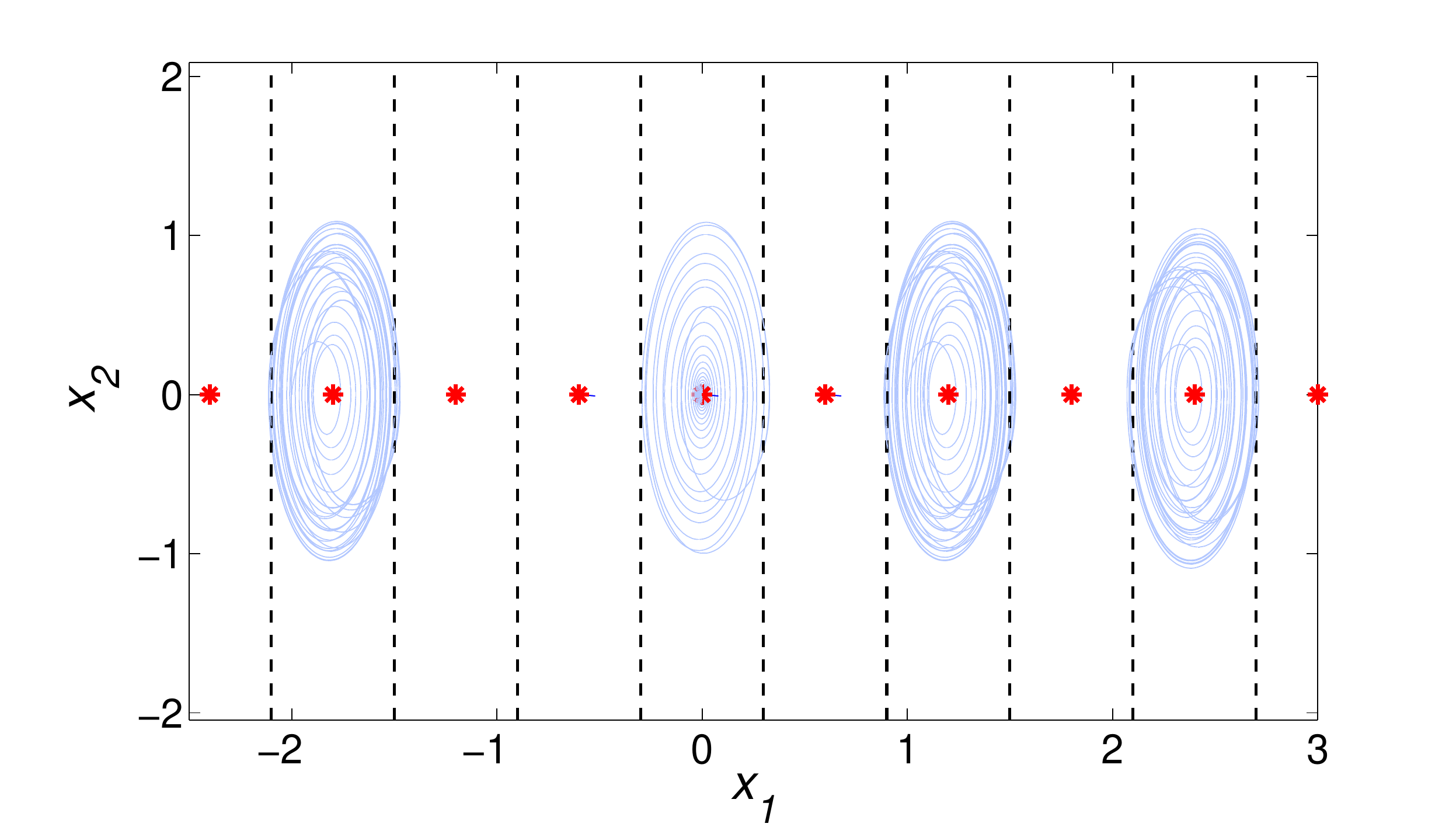}
\caption{\label{fig:multistability_2} Projections of the trajectories of the system  \eqref{ec_multistable} with \eqref{ec_round} onto the $(x_1,x_2)$ plane with $\nu=1.42$, $c=6.3$ and $\alpha=0.6$ for the initial condition sets $\mathbf{X}_0=(-2.0,0,0)^T$,  $\mathbf{X}_0=(0.01,0,0)^T$,  $\mathbf{X}_0=(1.0,0,0)^T$ and $\mathbf{X}_0=(2.21,0,0)^T$.}
\end{figure}

This comes as a mayor advantage in the proposed system as one may consider different initial conditions for values of $1.1<\nu<2.1$, and for each initial location given the system will oscillate near those continuous domains along the $x_1$ axis from $(-\infty,\infty)$. Consider the different sets of initial conditions   $\mathbf{X}_0=(-2,0,0)^T$,  $\mathbf{X}_0=(0.01,0,0)^T$,  $\mathbf{X}_0=(1.0,0,0)^T$ and $\mathbf{X}_0=(2.21,0,0)^T$ for $\nu = 1.42$. The resulting trajectory of the system given these conditions are depicted in Figure \ref{fig:multistability_2}, each scroll presented is a unique experiment for a given number of iterations in time. Notice that the orbit doesn't form attractors on the neighborhood domains. Resulting in a multistable system for the $x_1$ axis from $(-\infty,\infty)$ which oscillates around the equilibria in which the initial condition is set, and the trajectory is bounded by the dynamics of the left and right equilibrium points although there are no oscillations around them. Additionally,  the largest Lyapunov exponent of the attractor was calculated throughout the algorithm proposed by Rosenstein {\it et. al.} in \cite{rosenstein}, taking a value of $MLE=0.113669$ demonstrating a chaotic behaviour in the system.

%

\section{\label{sec:Conclusions}Concluding remarks}

Using the  Nearest Integer or $round(x)$ function it has been proved that multiple final states can be acquired in PWL systems considered in the  UDS of the type I theory.
The multistability phenomena is acquired by  means of the changes in the bifurcation parameter $\nu$, which correlate directly with the direction and location of their stable and unstable manifolds as well as the values of their eingenvalues.  It has been proven that with  specific $\nu$ values  the system may present multi-scroll attractors if the eigenspectra  satisfies the specifications discussed here. Otherwise the systems trajectory will be trapped between the adjacent equilibria regarding on the initial condition given to the system. Thus resulting in a multistable system along the $x_1$ axis.
The methodology discussed here may be applied to further axes in order to obtain multistability along  $\R^2$ or $\R^3$. The results of this may be reported elsewhere.

\section{Acknowledgements}
H.E.G.V is a doctoral fellow of the CONACYT in the Graduate Program on control and dynamical systems at DMAp-IPICYT. L.J.O.G. acknowledges the  UASLP for the financial support through  C16-FAI-09-46.46. E.C.C. acknowledges the CONACYT financial support for sabbatical at Department of Mathematics, University of Houston.


\end{document}